\documentclass[preprint,5p]{elsarticle}

\usepackage{amssymb}
\usepackage{amsmath}
\usepackage{lineno}

\biboptions{sort&compress}

\newcommand{\un}[1]{\ensuremath{\, \mathrm{#1}}}

\newlength{\delimw}
\newlength{\rxfc} 

\newcounter{temp}

\newcount\vone \newcount\vtwo \newcount\vthree \newcount\vfour \newcount\vfive  
\newcount\vsix \newcount\vseven \newcount\veight \newcount\vnine \newcount\vten 
\newbox\wbox \newdimen\wbsize                                                   
\def\underwiggle#1{\ifmmode\setbox\wbox=\hbox{$#1$}                             
                      \else\setbox\wbox=\hbox{#1}\fi                            
\dp\wbox=0pt\wbsize=\wd\wbox\lower2pt\hbox to0pt                                
{\hss$\vone=0\vtwo=0\vthree=7000\vfive=\vtwo                                    
\loop                                                                           
\vseven=\vone \divide\vseven by 2                                               
\vsix=\vfive \divide\vsix by 2 \multiply\vsix by -1                             
\veight=16384 \advance\veight by \vsix                                          
\vnine=16384 \advance\vnine by -\vsix                                           
\vten=\vseven \advance\vten by 32768                                            
\hskip\vseven sp                                                                
\vrule height\veight sp width 32768 sp depth\vnine sp                           
\hskip-\vten sp                                                                 
\ifdim\vseven sp<\wbsize \advance\vone by 20000                                 
\advance\vtwo by \vthree                                                        
\vfour=-\vtwo \divide\vfour by 10                                               
\advance\vthree by \vfour                                                       
\advance\vfive by \vtwo                                                         
\repeat$}\box\wbox}                                                             

\newlength{\figwidth}
\newlength{\figsep}
\newlength{\pba} \newlength{\wa} %Width of parbox and figure A
\newlength{\pbb} \newlength{\wb} %Width of parbox and figure B
\newlength{\pbc} \newlength{\wc} %Width of parbox and figure C
\newlength{\ha} %height of figure A

\newlength{\thinlinewidth} 
\newlength{\thicklinewidth} 

\usepackage{units}
\usepackage{booktabs} % for much better looking tables: \toprule, \midrule, \bottomrule
\usepackage{multirow}
\usepackage{xcolor} % guess what: colors!
\usepackage[final]{pdfcomment} % for comments: \pdfcomment, \pdfmargincomment, ...
\usepackage{amstext} % required for the \text command
\usepackage{subfig} 				% the subfigure package became obsolete, now use subfig with \subfloat[caption]{\includegraphics...}
\usepackage{arydshln}

\journal{Astroparticle Physics}

\begin{document}

\defineavatar{SB}{author=Sebastian,color=green}
\defineavatar{MK}{author=Marek,color=red}
\defineavatar{MV}{author=Markus,color=blue}
\defineavatar{LS}{author=Lukas, color=orange}
\defineavatar{NS}{author=Nora, color=yellow}
\listofpdfcomments[liststyle=AuthorComment]{}

\begin{frontmatter}



\title{Detecting extra-galactic supernova neutrinos in the Antarctic ice}


\author{Sebastian B\"oser\corref{au:sboeser}}
\author{Marek Kowalski}
\author{Lukas Schulte}
\author{Nora Linn Strotjohann}
\author{Markus Voge}
\cortext[au:sboeser]{Email: sboeser@physik.uni-bonn.de}

\address{Physikalisches Institut, Universit\"at Bonn, D-53115 Bonn, Germany}

\begin{abstract}
Building on the technological success of the IceCube neutrino
telescope, we outline a prospective low-energy extension that utilizes
the clear ice of the South Pole. Aiming at a $10\un{Mton}$ effective
volume and a $10\un{MeV}$ threshold, the detector would provide
sufficient sensitivity to detect neutrino bursts from core-collapse
supernovae (SNe) in nearby galaxies. The detector geometry and
required density of instrumentation are discussed along with the
requirements to control the various sources of background, such as solar
neutrinos. In particular, the suppression of spallation events induced by
atmospheric muons poses a challenge that will need to be addressed.
%For SNe, the resulting detector would reach beyond our own
%Galaxy, delivering between 8.3 and 33.3 regular core-collapse SN detections per decade.
Assuming this background can be controlled, we find that the resulting detector
will be able to detect SNe from beyond $10\un{Mpc}$, delivering between 10 and
41 regular core-collapse SN detections per decade. It would further allow to
study more speculative phenomena, such as optically dark (failed) SNe, where the
collapse proceeds directly to a black hole,
%, a phenomenon that could be detected
at a detection rate similar to that of regular SNe.
We find that the biggest technological
challenge lies in the required number of large area photo-sensors, with
simultaneous strict limits on the allowed noise rates. If both can be realized,
the detector concept we present will reach the required sensitivity
with a comparatively small construction effort and hence offers a
route to future routine observations of SNe with neutrinos.
\end{abstract}


\end{frontmatter}

%%\linenumbers

\section{Introduction}
\label{sec:intro}
%==============================================================================

In 1987, Supernova 1987A (SN 1987A) exploded in the Large Magellanic Cloud at a
distance of only 50\un{kpc}, leading to the first detection of neutrinos from
outside our solar system. Despite the fact that only $\sim$20 supernova
neutrinos were detected in total~\cite{Arnett}, a wealth of papers has been
published in its wake (see e.g.~\cite{GiuntiKim} for a summary), reflecting the
numerous and fundamental roles that neutrinos play in astrophysics as well as
in particle physics (e.g.~see~\cite{raffelt, dighe}).

Given today's detectors, a supernova in our Galaxy would result in $\sim 10^4$ neutrino
events detected individually in Super-Kamiokande~\cite{SuperK-SN} and other large low-energy
neutrino detectors, as well as up to millions of neutrinos detected through an increase in
noise rate in IceCube~\cite{IceCubeSN}. However, with an expected rate of only 1-3 Galactic SNe
per century~\cite{single_neutrino,deeptitand}, the chance for a detection during the lifetime
of the experiments is not overwhelmingly large. In the fortunate case of a SN detection, the
uniqueness of the progenitor system will make it difficult to disentangle the astrophysical
diversity from the effects due to particle physics (e.g.~neutrino oscillations) that will impact
the light curve and energy spectra.\\

As pointed out in~\cite{single_neutrino} and~\cite{deeptitand}, the situation
will change drastically once neutrino detectors reach the sensitivity threshold
to detect ``mini-bursts'' of neutrinos from supernovae in neighboring galaxies.
Not bound to our own Galaxy, the rate of SN observations will depend only on
the size of the detectors. As we will show in Section~\ref{sec:snrate}, an
effective volume of $\sim10\un{Mtons}$ is sufficient to detect SNe at a rate of
%$\sim3-5\un{yr^{-1}}$, % where did we get this from?
$\sim1-4$ per year---albeit most of them with less than ten individual neutrino
events. Despite the low number of detected neutrinos, these routine
observations would provide a wealth of information and allow entirely new
studies~\cite{deeptitand}. What follows is a brief and incomplete summary of
the scientific benefits of a large supernova neutrino detector.

The total SN rate in our local universe would be determined in a novel and less
biased way. Given an apparent mismatch in rates---only about half the expected
rate of SNe is actually detected by optical surveys~\cite{SNRfromSFR}---a
direct measurement would solve the riddle of missing supernovae.

Furthermore, a sufficiently sensitive detector allows us to test models
predicting additional neutrino bursts, such as failed or {\it dark}\/
(i.e.~optically unobservable) SNe~\cite{failed_sn}, merger of binary neutron
stars~\cite{merger} or the formation of quark stars~\cite{Dasgupta:2009yj}.
%An example of the diagnostic power provided by neutrinos
Dark SNe are core-collapse objects that directly form a black hole (BH).
Electromagnetic radiation is strongly suppressed, since the photons don't have
time to escape and are swallowed by the forming BH\@. Neutrinos, on the other
hand, can escape, and the expected burst from such an event is both more
luminous and hotter~\cite{failed_sn}. Average neutrino energies can be roughly
twice as large as in the case of the collapse to a neutron star (NS) and hence
open the opportunity to identify the collapse to a BH. % when it occurs.
%\pdfcomment[avatar=LS]{"identify the collapse to a BH whenever it occurs" mit
%weniger als 10 Neutrinos? Ist das nicht ein bisschen zu selbstbewusst?}
Another opportunity is the identification of questionable optical SN candidates,
e.g.~luminous blue variables~\cite{LBV}, as ``supernova impostors'' by the
non-detection of neutrinos~\cite{deeptitand}.

In addition, detections of neutrino bursts can be used to trigger early
optical or X-ray observations. Having a precise timing for the moment of
explosion and observing the shock-breakout in electromagnetic radiation, will
allow to infer a wealth of information about the progenitor system. In the
absence of a direction for the SN, the follow-up could focus on observation of
nearest galaxies, since these are the only ones for which a neutrino detection is
expected.

%For gravitational wave (GW) detectors, providing a time of explosion can reduce the background can be reduced by a factor of
%$\sim 10^7$ \pdfcomment[avatar=NS]{Ref?}. In the limit of large Gaussian backgrounds, the gain
%in This enhanced sensitivity would put the SNe detected by neutrinos
%within the reach of future GW detectors~\cite{Ott2008}.

With a larger number of supernova neutrinos, even a broader physics program
can be accessed. Due to the complexity of the involved processes, modeling of
supernova explosions is still a challenge today and has significant variance in e.g.~the
predicted mean neutrino energy~\cite{sato2}. Determination of the neutrino luminosity
and energy spectrum will provide valuable input to these models.

From the observation of the arrival times of the neutrinos from SN 1987A, a
limit in the eV range
%of $m_\nu < 5.8\un{eV}$
has been set on the effective mass of the anti-electron
neutrino~\cite{pagliaroli}. By observing neutrinos from supernovae with higher
statistics, more stringent limits could be set. In addition, by its impact on
the predicted flux, the neutrino mass hierarchy can be addressed as well, provided a
sufficient number of neutrinos is observed~\cite{IceCubeSN}.\\

Motivated by this scientific potential, several megaton scale neutrino
detectors are currently planned (e.g.~DeepTITAND~\cite{deeptitand},
Hyper-Kamiokande~\cite{hyperkamiokande,hyperk-loi} and UNO~\cite{uno}). Those
are either water Cherenkov detectors located in mines or marine detectors,
similar to ANTARES~\cite{Creusot:2013bwa}. In this paper, we explore the
potential of a $\sim10\un{Mton}$ detector in the Antarctic ice shield. In the
existing IceCube detector that has been optimized for TeV-PeV
energies~\cite{icecube}, SN neutrino bursts are typically searched for by
looking for a collective enhancement of photomultiplier noise
rates~\cite{IceCubeSN}. Due to the large sensor spacing and consequently high
energy threshold, attempting to detect individual neutrino events in this
configuration significantly reduces the effective mass and hence distance at
which SNe can be detected~\cite{Ribordy:1937}. A dedicated effort is now under
way to reduce the energy threshold to a few GeV in the PINGU (Phased IceCube
Next-Generation Upgrade) project~\cite{PINGU-LoI}, with one of its goals being
the determination of the neutrino mass hierarchy through the MSW
effect~\cite{Akhmedov:2012lr}. Building on this effort and the expertise
accumulated with IceCube and
AMANDA~\cite{Abbasi:2010qv,Achterberg:2006md,Collaboration:2011ym,Aartsen:2013jdh},
we explore the capabilities of a dedicated low-energy extension to study
individual supernova neutrino events. The challenge is to reduce the energy
threshold of the experiment by three orders of magnitude while controlling the
background at a level required for the detection of supernovae.

While we will focus on this aspect in this paper, it should be noted that such a
detector will not be limited to the detection of neutrinos at a few MeV, but
will also provide unprecedented
sensitivity at the GeV-scale. Not only the neutrino oscillation sector will be
accessible with increased precision --- it will also provide sensitivity
to other astrophysical phenomena such as collisional heating in gamma-ray
bursts~\cite{Bartos:2013hf,Murase:2013hh}.\\

The paper is organized as follows: The SN models that we use to benchmark a
future detector are discussed in Section~\ref{sec:ccsn}, followed by a
description of our detector simulation
%(Section \ref{sec:sim})
and the optimization of the detector configuration in Section~\ref{sec:simopt}.
Section~\ref{sec:background} is a discussion of the various background sources
that can be anticipated while Section~\ref{sec:diffusive} investigates an
alternative detector depth. Finally, we give an estimate on the expected SN rate
in Section~\ref{sec:snrate} and conclude with a discussion of the results in
Section~\ref{sec:conclude}.

%==============================================================================
\section{Neutrinos from core-collapse supernovae}
\label{sec:ccsn}
%==============================================================================
\begin{figure}
    \includegraphics[width=\linewidth]{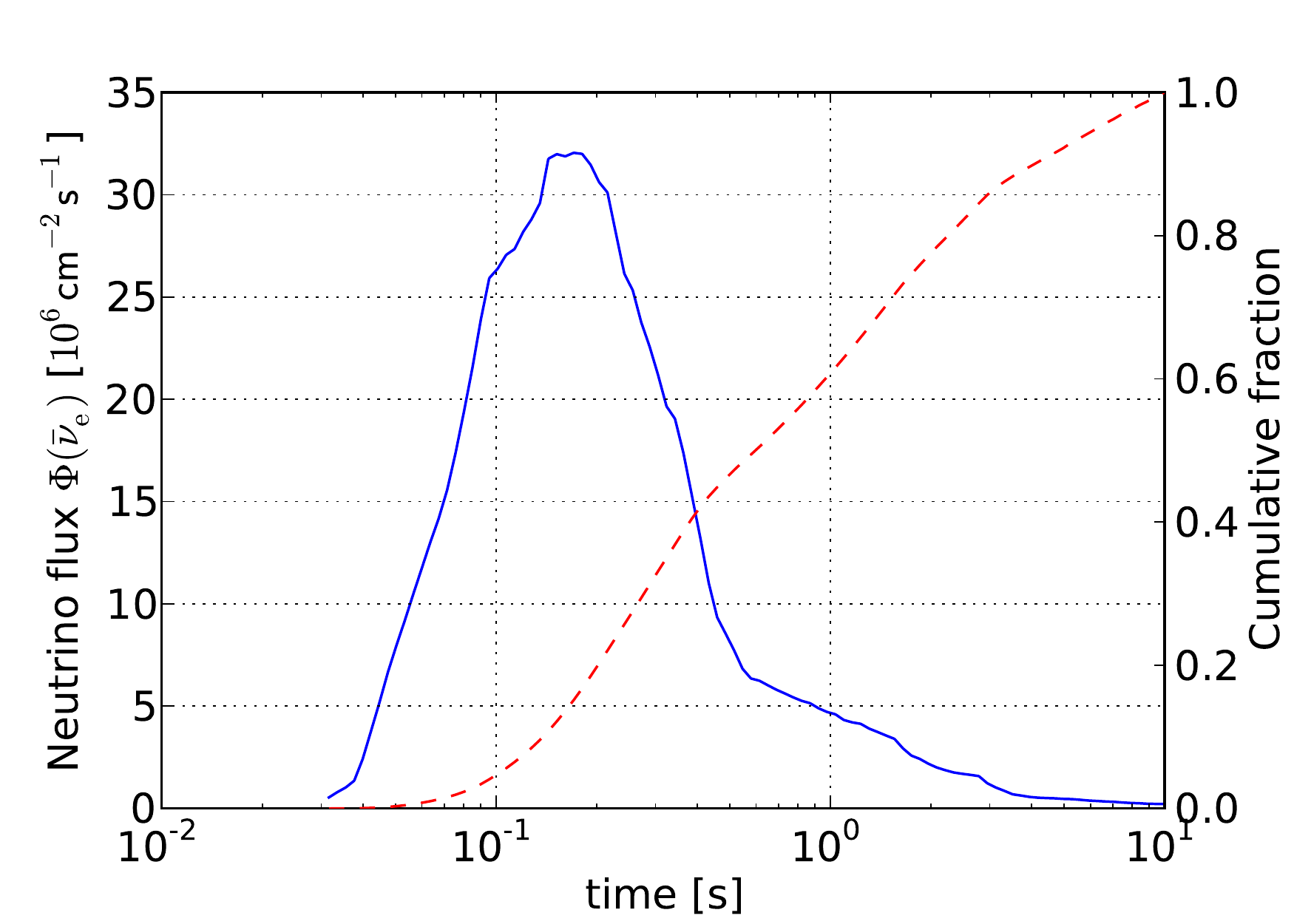}
    \caption
    {Time evolution of $\bar\nu_e$ flux for a SN at $1\un{Mpc}$ in the
    Lawrence-Livermore model~\cite{livermore} (full line
    represents the differential flux and dashed line the integral).\label{fig:nuflux}}
\end{figure}

In core-collapse supernovae, neutrinos are produced in course of the formation of a neutron star
which follows the gravitational collapse of a massive progenitor star. Electrons and protons
rapidly fuse to neutrons, emitting electron neutrinos, which results in a sharp ($\sim10\un{ms}$)
deleptonization peak in the neutrino flux which marks the birth of the neutron star. However, most
supernova neutrinos ($\sim90\%$) originate from thermal pair creation~\cite{neutrinophysik}. These
neutrinos, released over a characteristic diffusion time of 1-10\un{s}
(cf.~Fig.~\ref{fig:nuflux}), are of all flavors, with flavor ratios, light
curve and energy spectrum still being under debate~\cite{sato2}.

An early simulation of neutrino production in SN explosions has been provided in the so-called
Lawrence Livermore (LL) model~\cite{livermore}. In the LL model, the neutrino spectrum is
parametrized by
\begin{equation}
\label{eqn:nuenergy}
\frac{d N}{d E}=\frac{(1+\beta)^{1+\beta} L}{\Gamma(1+\beta) \langle E \rangle
^2} \left( \frac{E}{\langle E \rangle}\right)^\beta \exp{\left(
-(1+\beta)\frac{E}{\langle E \rangle} \right)}\mathrm{,}
\end{equation}
where $L$ is the luminosity and $\langle E \rangle$ the average energy while $\beta$ determines
the width of the spectrum.
%The parameters of the model can differ for each neutrino flavor.
The parameters for the LL simulation as well as for a competing model by Thompson, Burrows and
Pinto (TBP)~\cite{tbp_sn} are reproduced in Table~\ref{tab:sn}. While the latter model %from TBP
does not result in an explosion of the supernova,~\cite{sato2} suggests that the resulting
luminosity should be of the same order of magnitude. A large variety of other
neutrino emission models exists~\cite{Janka:2012wk}, but we restrict ourselves to these two commonly used ones for ease of
comparison.
%%The energy spectrum for electron antineutrinos is plotted in figure (\ref{fig:neutrino_spectrum}).
\begin{table}[tbp]
  \begin{center}
    \begin{tabular}{rcccc}
      \toprule
      & Mass [$M_\odot$] 	& $\langle E_{\overline{\nu}_e} \rangle$ [$\unit[]{MeV}$] &
      $\beta_{\overline{\nu}_e}$ & $L_{\overline{\nu}_e}$ [$\unit[]{erg}$] \\
      \midrule
      LL SN & $20$ & $15.4$ & $3.8$ & $4.9 \cdot 10^{52}$ \\
      TBP SN  &  $11$ and $15$  & $11.4$ & $3.7$ & ($4.9 \cdot 10^{52}$) \\
      \hdashline[2pt/5pt]
      Dark SN  &  $25 - 40$ & $20 - 24$ & - & $ \sim10^{53}$ \\
      \bottomrule
    \end{tabular}
    \caption[Parameters of the LL, the TBP and dark supernova model]{Parameters of the Lawrence
    Livermore (LL) model~\cite{GiuntiKim}, the Thompson, Burrows, Pinto (TBP) model~\cite{tbp_sn}
    and the dark supernova model~\cite{failed_sn} for the neutrino spectrum of a core-collapse
    supernova. The table presents the stellar mass of the progenitor $M_\odot$, the average neutrino
    energy $\langle E_{\overline{\nu}_e}\rangle$, the pinch parameter $\beta_{\overline{\nu}_e}$ (see Eqn.~\ref{eqn:nuenergy}) and the time-integrated
    luminosity in anti-electron neutrinos $L_{\overline{\nu}_e}$. The TBP model does not lead to an explosion, i.e.\
    no luminosity emerges from the simulation. Instead, the luminosity from the LL model is
    assumed.  Please note that for dark SNe Eqn.~\ref{eqn:nuenergy} is not
    valid, instead the positron spectrum given in~\cite{failed_sn} was used.
    %\pdfcomment[avatar=LS]{Wo kommt die Zahl in Klammern (wenn sie mal da steht) denn dann her?}
    }
    \label{tab:sn}
  \end{center}
\end{table}
Alternative models predicting low-energy neutrinos include {\it dark
supernovae} \/in which stars heavier than $\sim \unit[25]{M_\odot}$ form black
holes. It is assumed that if the progenitor does not rotate fast enough to
explode as a hypernova, it will be very faint or even dark in optical emission,
while even more luminous in neutrino flux than ordinary
supernovae~\cite{failed_theo, Sumiyoshi:2006id}. The resulting neutrinos have
average energies of $\left<E_\nu\right> \sim 20-24\un{MeV}$~\cite{failed_sn}.
\begin{figure}
\includegraphics[width=\linewidth]{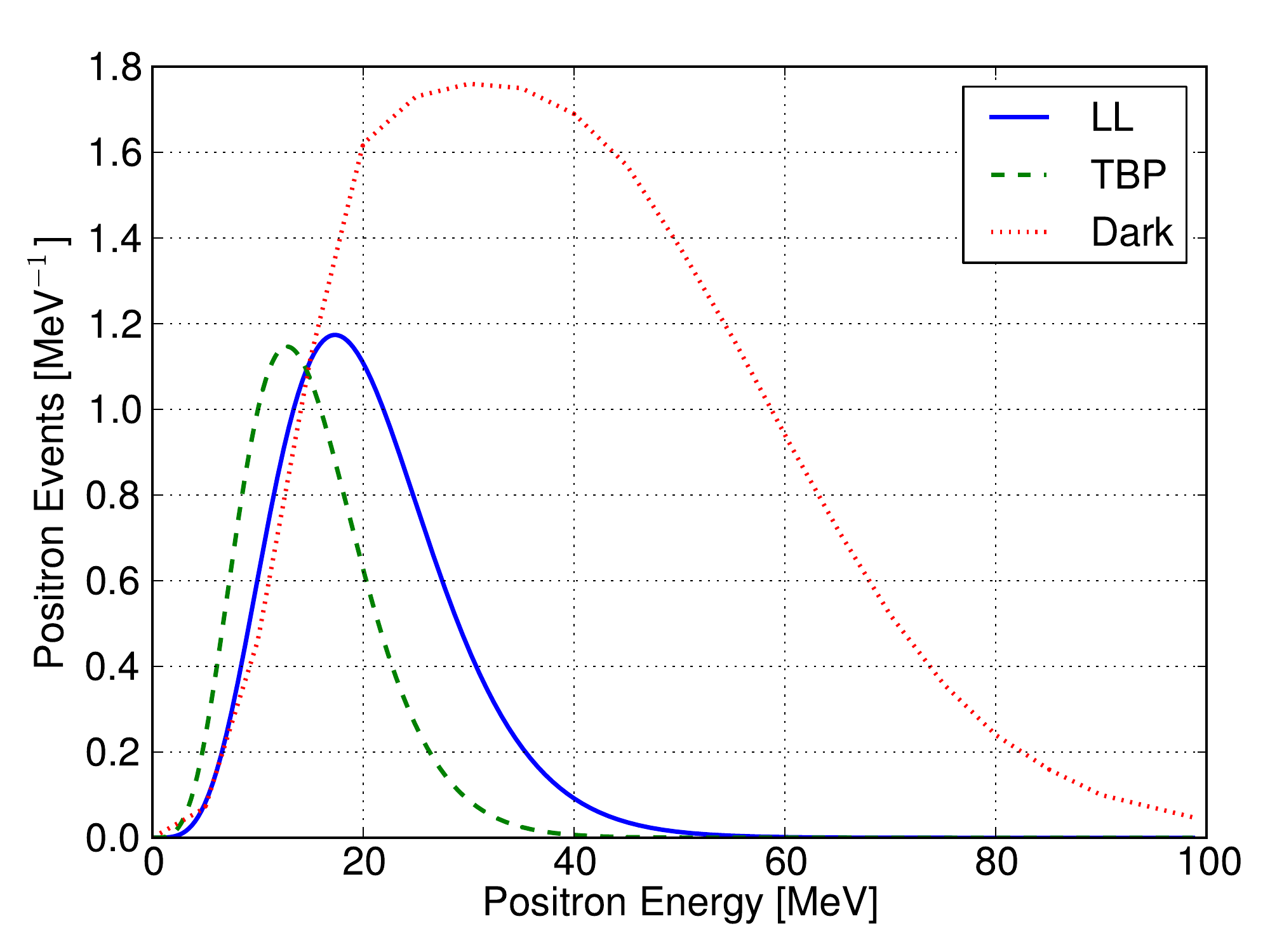}
\caption
{Positron spectrum for the Lawrence-Livermore (LL)~\cite{livermore}, Thompson, Burrows, Pinto (TBP)~\cite{tbp_sn}
 and a dark supernova model~\cite{failed_sn} for a SN at $1\un{Mpc}$
 and an effective detector volume of $1\un{Mton}$.\label{fig:positron-spec}
}
\end{figure}

The neutrinos from the different collapse scenarios can best be detected via inverse beta decay (IBD)
\begin{equation}
    \overline{\nu}_e + p \rightarrow e^+ + n
\end{equation}
that requires a threshold energy of $E_\nu > 1.806\un{MeV}$. We use the approximation for the energy dependence of the cross-section called ``Na\"ive +``, which is presented in~\cite{strumia}.
%precise approximation of
%According to~\cite{strumia}, a
%precise approximation of the cross-section up to energies of $300\un{MeV}$ is given by:
%\begin{equation}
 %   \sigma(\overline{\nu}_e p)\approx \unit[10^{-43}]{cm^2} p_e E_e E_\nu^{a+b\ln{E_\nu}-c\ln^3{E_\nu}} \mathrm{, } \qquad \mathrm{with } E_e=E_\nu-\unit[1.293]{MeV}
%    \label{eqn:inv-beta}
%\end{equation}
%where $p_e$ is the momentum of the resulting positron, $E_e$ its
%energy and $E_\nu$ the energy of the neutrino in $\unit[]{MeV}$. The numeric
%values of the parameters are $a=-0.07056$, $b=0.02018$ and $c=-0.001953$.
Since the cross-section rises with energy, the resulting positron spectrum is harder than the
initial neutrino spectrum. Figure~\ref{fig:positron-spec} shows the resulting positron spectra
for the two different supernova models as well as for a model of dark
supernovae. Neutrino oscillations, which harden the spectrum, are taken into
account only for the dark model, for which we directly take the positron
spectrum from~\cite{failed_sn}.

%\begin{figure}[tbp]
%  \begin{center}
%    %\includegraphics[width=0.5\textwidth]{positron_spectra_color}
%    \includegraphics[width=0.5\textwidth]{positronSpectra_1Mpc_1Mton}
%  \end{center}
%  \caption[Positron spectrum of supernova models]{The positron spectrum for three supernova
%  models: Lawrence-Livermore (LL)~\cite{livermore}, Thompson, Burrows, Pinto (TBP)~\cite{tbp_sn}
%  and a dark supernova model~\cite{failed_sn}. The SN takes place at a distance of $1\un{Mpc}$
%  and the effective detector volume is $1\un{Mton}$. Neutrino oscillations, which harden the
%  spectrum, are taken into account only for the dark model.}
%  \label{fig:positron-spec}
%\end{figure}
For a typical positron energy of $20\un{MeV}$, the corresponding track length
is $\sim10\un{cm}$ in ice, resulting in $\sim3600$ Cherenkov photons ($300-600\un{nm}$)~\cite{IceCubeSN}.
Since the light yield scales linearly with the positron track length, and hence
with the positron energy, the average amount of light produced per neutrino is
model dependent.

%==============================================================================
\section{Detector simulation and optimization}
\label{sec:simopt}
%==============================================================================

\begin{figure}[tb]
    \begin{center}
        \includegraphics[width=0.9\linewidth]{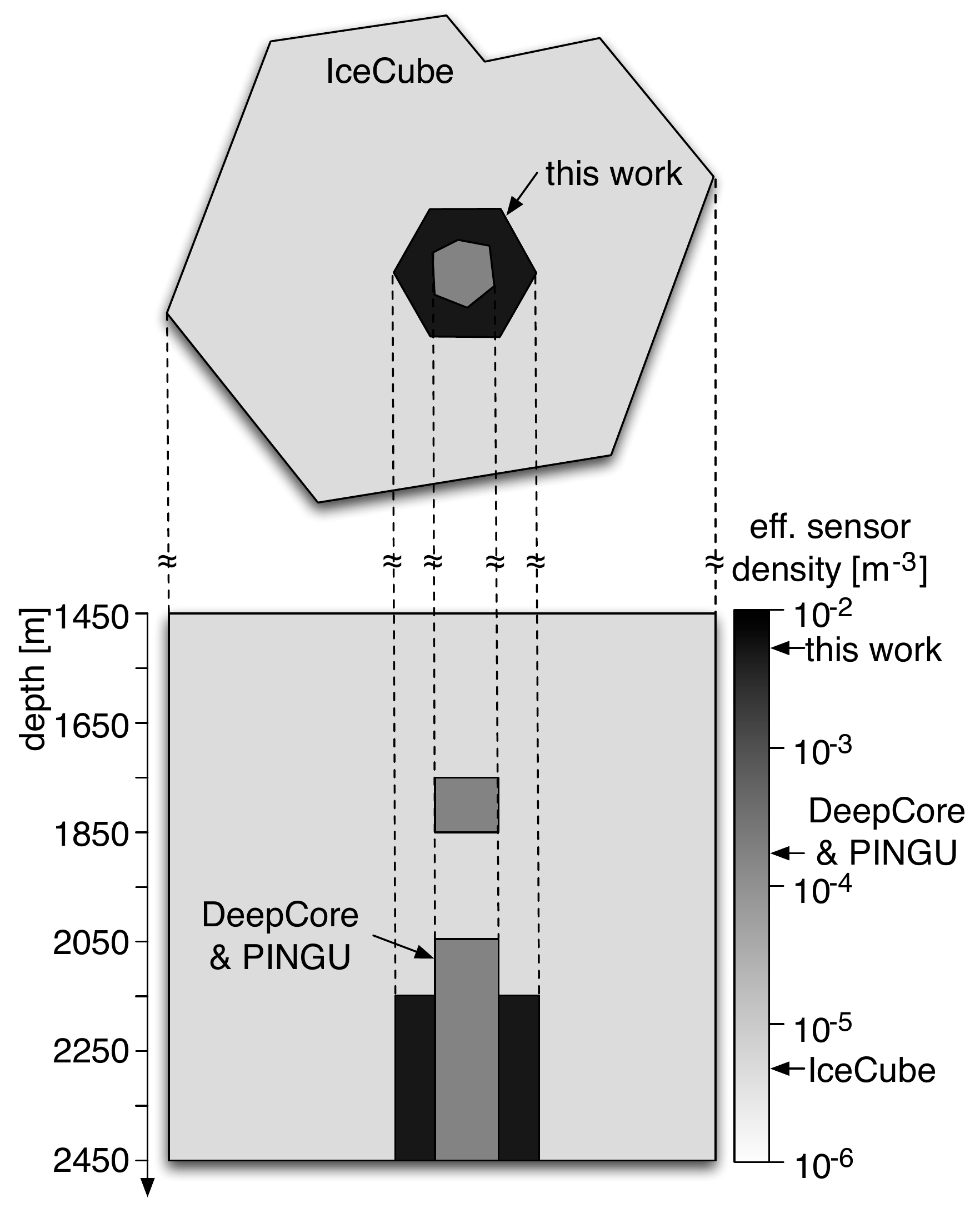}
    \end{center}
    \caption
    {Dimensions and effective sensor density of the aspired detector
    %of the final detector layout in the clear ice
    with 127 strings and 25\un{m} string spacing as compared to
    IceCube and PINGU/DeepCore. The gray-scale effective sensor
    density indicates the installed photo-effective area per
    $\un{m^3}$ in units of HQE IceCube DOMs.\label{fig:geo}}
\end{figure}

In this section we describe the simulation and optimization of a possible
optical Cherenkov detector array in the Antarctic ice, capable of detecting MeV
neutrinos with high statistics.  The geometry of the detector as sketched in
Figure~\ref{fig:geo} consists of 127 vertical strings, arranged in a filled
hexagon, similar to IceCube and corresponding to a construction period of $\sim
5\un{years}$. Fuel costs dominate the drilling expenses which amount to about
$0.5 {\rm M\$}$ per hole. While the number of strings is comparable, closer horizontal
spacing of the strings than in IceCube is required. We optimize this spacing between 10 and
$\unit[40]{m}$ for the best effective detection volume. For the vertical position
of the optical sensors we choose the ice layer between $2150$ and  $2450\un{m}$
below the surface, where also the DeepCore array is located. At this depth, air
bubbles have fully degenerated due to the high ambient pressure and only a
small dust concentration was measured~\cite{icepaper,Aartsen:2013uq}.
Consequently, the ice in these depths has the largest scattering lengths of
$\lambda_e \approx 20-50\un{m}$. The absorption length is $\lambda_a \approx
20-90\un{m}$, depending on wavelength~\cite{icepaper,Aartsen:2013uq}.

These excellent optical properties---comparable to what can be achieved in
laboratory conditions---allow for an efficient photon detection as well as
precise reconstruction of the event position and direction. For an accurate
simulation of the photon propagation in this environment, we use the {\it
Photonics code}~\cite{photonics, photonics-web, photonics-code}---a parametric
simulation developed for IceCube.  It includes the full depth-dependence of the
scattering and absorption properties of the ice~\cite{icepaper}.

IceCube employs photo-sensors,~so-called digital optical modules
(DOMs)~\cite{Abbasi:2008aa}, each including a 10'' Hamamatsu PMT
integrated into a pressure-resistant glass sphere that also includes
the electronics for HV generation and in-ice digitalization of the PMT
signal.  The dark noise rate of individual DOMs averages around
$500\un{Hz}$~\cite{IceCubeSN}. IceCube DOMs have a non-trivial
directional sensitivity~\cite{Abbasi:2010vc} which is incorporated in
the Photonics simulation package~\cite{photonics-code} and hence
included in our simulation.  Both regular efficiency and high quantum
efficiency (HQE) PMTs are deployed in IceCube. From laboratory
measurements~\cite{WOM-paper}, we obtain an effective area\footnote{Averaged over
all incident angles and wavelengths, assuming isotropic emission with
a Cherenkov spectrum $\propto\lambda^{-2}$.} of
$\unit[19.4 \, (26.3)]{cm^2}$ for an IceCube DOM
equipped with a regular (HQE) PMT.
% XXX: Explain that we simulate a module that has the area of a cylinder (as in shallow ice),
% but the angular acceptance of IceCube DOM

Due to the comparatively low energy of supernova neutrinos and the
corresponding small light yield, a high density of photo-sensitive
area is required to obtain acceptable trigger rates. We find that even
for the closest possible vertical spacing of one HQE optical module
per meter on our 300\un{m} long strings and the closest possible
horizontal string spacing of 10\un{m}, the effective mass for neutrino
detection falls short of our 10\un{Mton} target
(c.f.~Fig.~\ref{fig:effmass_pscut}). Instead, we simulate
photodetectors with an effective area
%corresponding to a cylinder of
%$1\un{cm}$ circumfence and a length of one meter (but with the same
%angular acceptance as IceCube DOMs),
equivalent to $\approx 5.4$ HQE
IceCube optical modules per meter. While the use of significantly
larger photocathode area may be challenging and is in particular
limited by the achievable drill hole diameter, on-going R\&D for
larger effective area and cheaper optical modules with less noise
based on wavelength shifters shows promising first
results\cite{WOM_ICRC}.\\

Neutrinos are simulated employing a Monte Carlo method
where interaction vertices are generated homogeneously within and beyond the
detector volume. The positron energy is sampled according to one of the energy
spectra shown in Fig.~\ref{fig:positron-spec}.
%\pdfcomment[avatar=NS]{Nee, die Positronenenergie wird aus der Neutrinoenergie
%berechnet.}
Using the Photonics code, we first calculate the average number of detected
photons for each sensor, given the neutrino vertex and positron energy.
%%%%% old for deep and shallow side-by-side %%%%%
%For the diffusive ice, this is done via analytic functions, for the clear ice,
%the Photonics code is employed.
The actual number of detected photons is then drawn from a Poisson distribution
and the hit times are sampled for each of these photons from the respective
arrival time distribution.

%%%%% old for deep and shallow side-by-side %%%%%
%% effectiveMass/plotEffectiveMassesVsSpacing.py
%\begin{figure}
%\includegraphics[width=\linewidth]{effectiveMass_forPaper_61strings_oldsim}
%\caption{The effective detector mass at trigger level for a 61-string geometry
%  (averaged over the LL model spectrum) as function of string spacing. In clear
%  ice (dashed) the effective mass is several times lower than in diffusive ice
%  (solid) due to ice properties. Only if the sensor effective area is increased
%  by a factor of four (dotted), similar results can be obtained.
%  \label{fig:eff-mass}}
%\end{figure}

In order to optimize the detector geometry, the effective positron detection
mass is calculated as function of the lateral spacing between strings.  The
\emph{effective mass} is the mass of the geometrical simulation volume
multiplied with the fraction of simulated events that are detected above
threshold.  This threshold is set at a minimum of five photo-sensors being hit
by photons, and was chosen to allow for reconstruction of the vertex position and
positron direction which correspond to five degrees of freedom.
%%%%% old for deep and shallow side-by-side %%%%%
%Figure~\ref{fig:eff-mass} shows the resulting effective mass as a function of the string spacing
%for the clear and diffusive ice. %For the clear ice we have explored higher photo-sensor densities as well.
Generally, one finds that, as the string distance increases, the geometrical
detector volume increases, but the fraction of detected neutrino events with at
least five hit sensors decreases, yielding a maximum effective volume of $\sim
18\un{Mton}$ for a string spacing around $30\un{m}$ for a Lawrence-Livermore
neutrino spectrum.
\section{Background studies}
\label{sec:background}
%==============================================================================

Up to this point, we have solely implied the trigger requirement---that the number of sensors hit by photons
be at least five for each neutrino event. In this section, we will discuss the dominant sources
of background and how they can be controlled by implying additional constraints
on the distribution of photon hits. Contributions from random noise, atmospheric
neutrinos, atmospheric muons, and solar neutrinos are considered.\\

Neutrinos from SNe come in bursts. To be distinguished from uncorrelated
background or noise triggers, a SN will need to produce a certain multiplicity
of neutrino triggers within a given time window. We can claim a supernova
detection if a certain number of trigger events $N_\nu$ occur within a time
window of $\Delta t_{SN} \sim 1-10\un{s}$ (c.f.~Figure~\ref{fig:nuflux}). Under
the constraint that we want a limited amount of false SN detections, $N_\nu$
and $\Delta t_{SN}$ determine the maximally allowed background/noise trigger
rate $f_\mathrm{noise}$. The number of false SN detections per year
$N_\mathrm{SN}$ as function of $f_\mathrm{noise}$, $N_\nu$ and $\Delta t_{SN}$
is:
\begin{equation}
    N_\mathrm{SN} = f_\mathrm{noise} \cdot \left(1 - P_\mathrm{cdf}(N_\nu-2, \mu=f_\mathrm{noise} \Delta t_{SN}) \right) \cdot \mbox{1 yr}
    \label{eqn:fnoise}
\end{equation}
where $P_\mathrm{cdf}$ is the cumulative Poisson distribution. If we
want to limit the false SN detection rate to about 1 per year,
comparable to the expected signal SN detection rate, we get the
maximally allowed noise rates shown in Table~\ref{tab:fnoise}. E.g.,
we can accept at most $\unit[\approx0.9]{mHz}$ of background/noise
trigger rate if we want to detect a SN with 3 neutrino events within
$\unit[10]{s}$. In the following estimates, we will use the rounded
value of $\unit[1]{mHz}$ as the tolerable upper limit on the noise and
background rates.

\begin{table}
    \begin{center}
        \begin{tabular}{ccc} \toprule
            $N_\nu$ & $\Delta t_{SN}$ [s] & $f_\mathrm{noise}$ [mHz] \\ \midrule
            %3 & 1 & 3.99 \\
            %4 & 1 & 20.96 \\
            %5 & 1 & 60.32 \\
            %6 & 1 & 127.17 \\ \midrule
            3 & 10 & 0.86 \\
            4 & 10 & 3.74 \\
            5 & 10 & 9.61 \\ \bottomrule
           % 6 & 10 & 18.82 \\ \bottomrule
        \end{tabular}
        \caption[Maximally allowed noise trigger rate]{Maximally allowed
        noise/background trigger rate $f_\mathrm{noise}$ for an average of 1 false
        SN event per year, consisting of $N_\nu$ events falling into a time
        window of $\Delta t_{SN}$.}
        \label{tab:fnoise}
    \end{center}
\end{table}

\subsection{Sensor noise}
\label{sec:noise}

Since the Antarctic ice shield is a very low-radioactivity
environment, the main sources of random noise are introduced by the
detector itself: radioactive isotopes and thermal noise in the
photo-sensors and electronics. For the IceCube modules, this results
in a dark count rate of $\sim 500\un{Hz}$~\cite{IceCubeSN}.  It is
known that some fraction of the noise is not purely random, but
correlated in time, however we will neglect this for the sake of
simplicity.
%As was shown above, the effective photosensitive area will have to be increased to about
%$0.014\un{m^2}$ per instrumented meter to achieve the effective target mass of $10\un{Mtons}$.
%While conventional PMTs of this size are available, the required thickness of a glass pressure
%vessel of this size and the cost to drill holes of the corresponding diameters make this option
%very unattractive.

As mentioned above, the detector presented here will not be feasible using IceCube
modules.  New photo-sensor technologies (e.g.\ based on
wavelength shifters as light collectors) are currently discussed for deployment
in future extensions to IceCube \cite{WOM_ICRC},
%\pdfcomment[avatar=SB]{Make a reference to future WOM paper?}
that offer increased effective photo-sensitive area in combination with a
significantly reduced noise rate. These technologies are
still in the design phase, so the achievable noise rate is not known yet.
We use dark noise values of $500\un{Hz}$, $50\un{Hz}$
and $10\un{Hz}$ as templates in absence of solid numbers.
%As the detector studied here has a much larger sensitive
%area per sensor, a noise rate of $500\un{Hz}$ per sensor will be assumed in the following.

As shown in Table~\ref{tab:fnoise}, the rate $f_\mathrm{noise}$ of background
neutrino triggers caused by random noise hits has to be below $\approx
1\un{mHz}$ if only one false 3-neutrino supernova burst detection is tolerated
per year. We calculate this noise trigger rate depending on the total
number of modules in the detector $N_{\mathrm{tot}}$, the random noise rate per
module $f_m$, the number of hit modules $n_{\mathrm{trig}}$ that is required
for a neutrino event to trigger and the trigger time window
$t_{\mathrm{trig}}$. Assuming that one module has registered a random noise
hit and opened the trigger window, the probability $P_m$ for any module to also
see at least one noise hit during this time window $t_{\mathrm{trig}}$ is
complementary to the probability to register no hit, which will follow a
Poisson distribution:
%
% The random noise has a Poisson distribution and can be calculated accordingly. So for a single
% sensor the probability
% to
% see at least one random noise hit during the time window $t_{\mathrm{trig}}$ is complementary to the probability to
% register no hit under the
% specified coincidence conditions and hence given by
\begin{equation}
P_m = 1 - e^{-f_m\cdot t_{\mathrm{trig}}}.
\end{equation}
The probability for a noise event $P_{\mathrm{noise}}$ is the probability that at least
$n_{\mathrm{trig}} - 1$ more modules also encounter a noise hit in the time
window. Using the binomial distribution, again via the
complementary probability of registering $n_{\mathrm{trig}} - 2$ noise
hits or less:
\begin{equation}
    P_{\mathrm{noise}} = 1 - B_{\mathrm{cum}}(n_{\mathrm{trig}} - 2 \,|\, N_{\mathrm{tot}}, P_m) \; ,
\end{equation}
where $B_{\mathrm{cum}}(m \,|\, n, p) = \sum_{k=0}^{m} {n \choose k} \; p^k \; (1-p)^{n-k}$ is the
cumulative binomial probability for up to $m$ successes out of $n$ tries with probability $p$.
Apart from boundary effects, the rate of noise triggers in the detector is then
\begin{align}
    f_{\mathrm{noise}} &= P_{\mathrm{noise}} \cdot f_m \cdot N_{\mathrm{tot}}.
\label{eqn:noise}
\end{align}

% \begin{figure}[tbp]
%   \begin{center}
%     \includegraphics[width=\textwidth]{hitTimeDistribution}
%   \end{center}
%   \caption[Time distribution of arriving hits]{Cumulative distribution of the arrival time of
%   the photon hits from triggered events (with at least 5 sensors hit) in the clear ice and diffusive
%   ice detector.}
%   \label{fig:coincidence}
% \end{figure}

% For $n_{\mathrm{trig}} = 5$ and demanding $f_{\mathrm{noise}} \leq 4\un{mHz}$, we can now adjust the coincidence window.
% As one can see from Figure \ref{fig:coincidence},
% %\pdfcomment[avatar=LS]{Fig. 3.11 aus Noras Arbeit übernehmen? Oder reicht es, wenn wir die Zeiten einfach angeben?}
% for the diffusive ice a time window of
% %$\sim 2500\un{ns}$ is required to
% $\sim 2300\un{ns}$ is required to
% catch a major fraction of the signal hits. According to Eqn.~\ref{eqn:noise} the spatial coincidence window may then
% %contain not more than 9 sensors,
% contain not more than 10 sensors,
% which in turn reduces the signal efficiency dramatically.
% In the clear ice, however, the amount of unscattered photons is much higher, significantly
% decreasing the mean delay time between the signal photons. Thus the coincidence time window can
% be set to only
% %$500\un{ns}$, allowing for up to 42 sensors in spatial coincidence.
% $1000\un{ns}$, allowing for up to 21 sensors in spatial coincidence.
% Additionally, one can make use of the hit pattern forming a Cherenkov cone when designing the
% spatial cut, whereas in the diffusive ice all geometrical information is destroyed and only a
% simple spherical cut is possible.

Using generic values for the dark count $f_m = 500\un{Hz}$,
$N_\mathrm{tot} = 38100$ for a detector
with 127 strings and 300 modules per string, and requiring $n_\mathrm{trig}=5$
hits in $t_\mathrm{trig}=1000\un{ns}$, the rate of false SN events is
$f_\mathrm{noise}=19\un{MHz}$, i.e. 10 orders of magnitude above the allowed
value from Table~\ref{tab:fnoise}. Even assuming a module dark noise rate as low as
$f_m = 10\un{Hz}$, still $f_\mathrm{noise} = 247\un{Hz}$, well in excess of what
can be tolerated. This clearly shows that it is necessary to apply intelligent
trigger algorithms that take advantage of the non-uniform distribution of
photons from neutrino interaction and thus limit the number of modules
considered by the trigger. In the following,
we present two strategies that reject events induced by sensor noise.

\subsubsection{Local coincidence cleaning (RT)\label{sec:rtcleaning}}

\begin{figure}[t!]
\includegraphics[width=\linewidth]{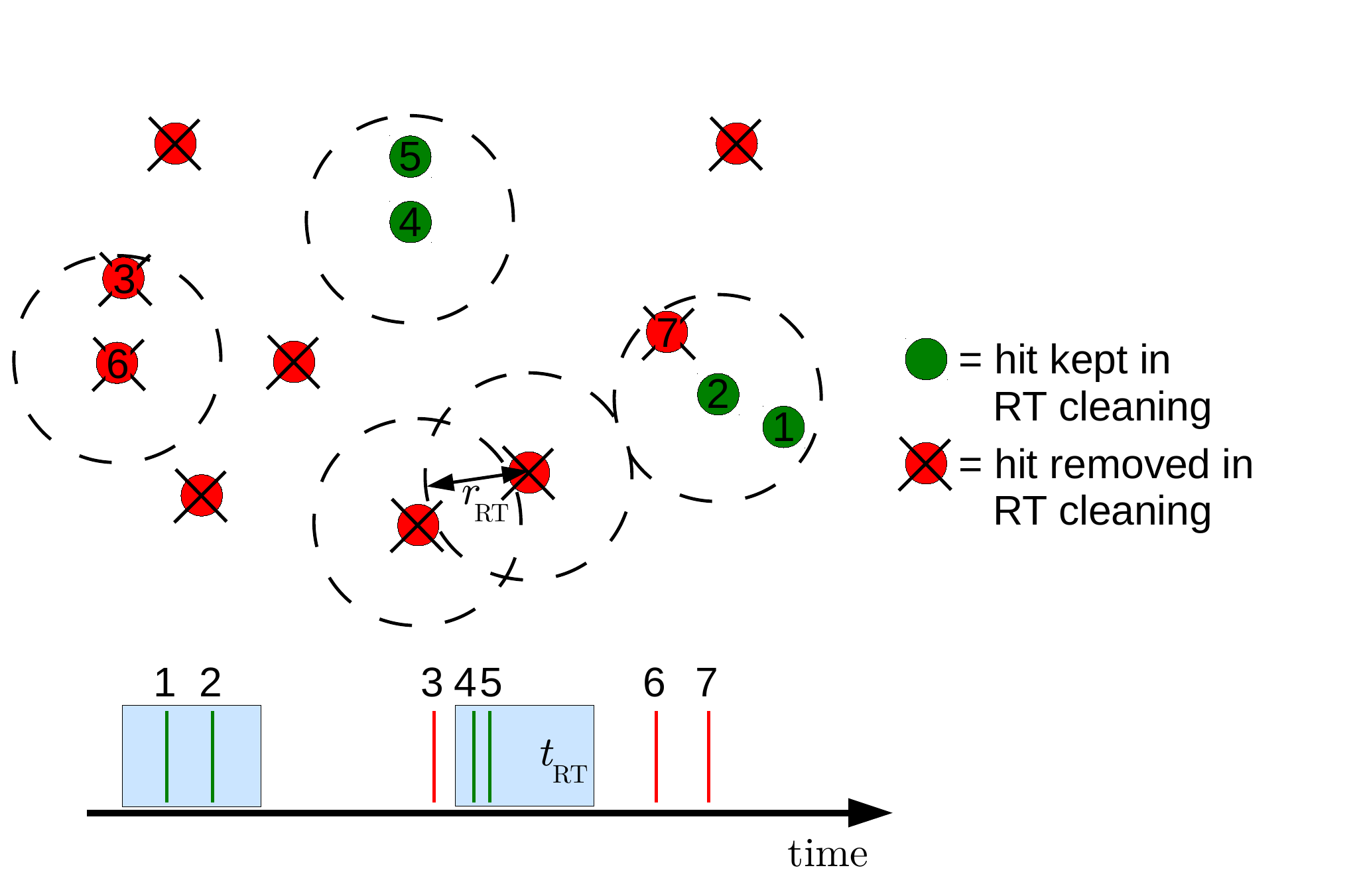}\\
\includegraphics[width=\linewidth]{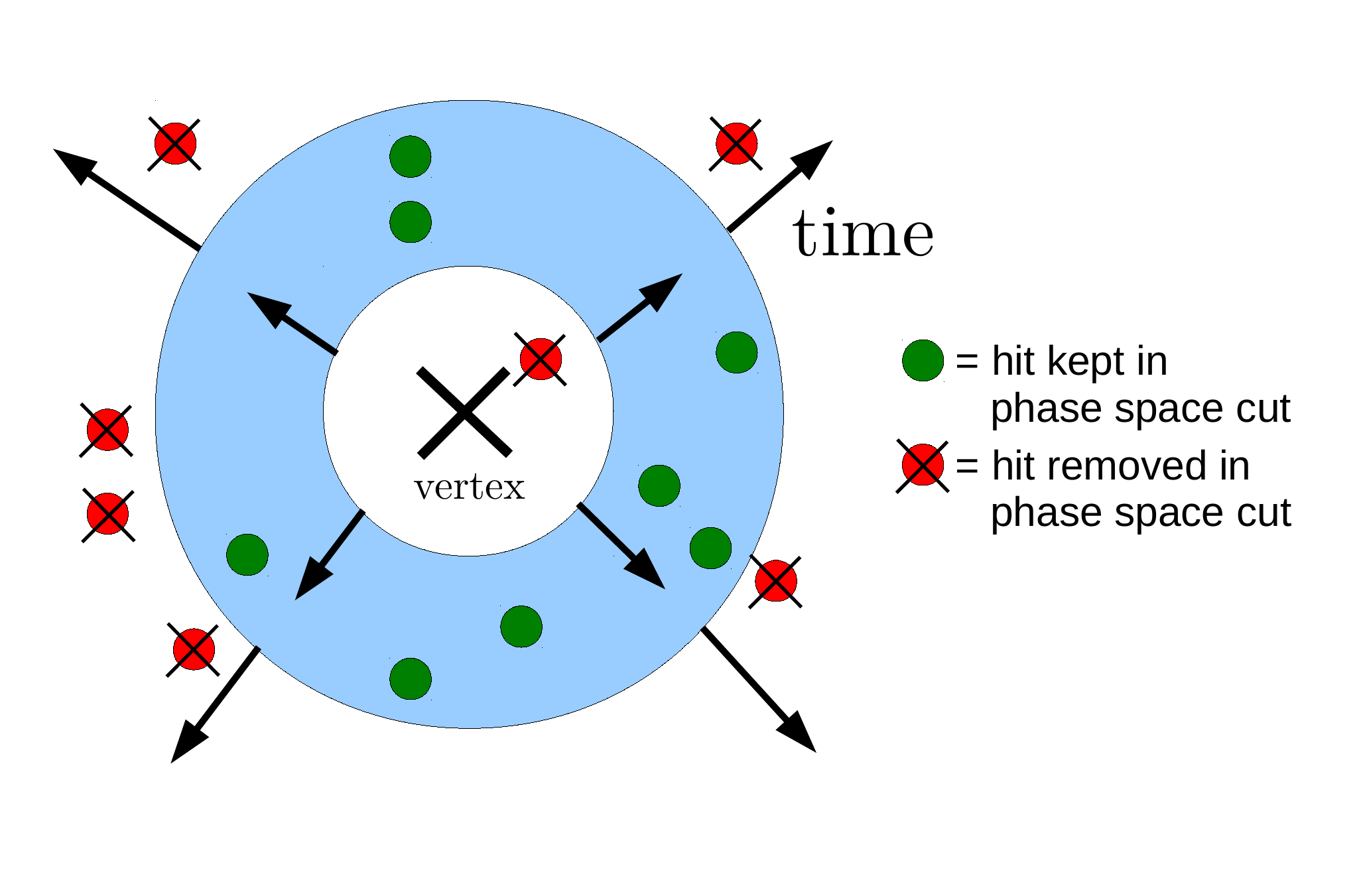}
    \caption{Illustration of the local coincidence (RT) cleaning and the phase
    space (PS) cut. Each colored sphere corresponds to a hit optical module.
    Dashed circles indicate the radius $r_\mathrm{RT}$, blue boxes the time
    window $t_\mathrm{RT}$ of the local coincidence cleaning. The blue region
    marks the expanding spherical shell used in the phase space
    cut.\label{fig:rtcleaning}}
\end{figure}

%While taking advantage of the localization of the event, the above method does
%not account for the fact that light from the positron is emitted in a preferred
%direction on the Cherenkov cone.

Noise hits are uniformly distributed accross the detector while signal hits
follow a certain topology: A positron from inverse beta decay produces
Cherenkov light along its few cm long track. We can consider the positrons to
be point sources of light that is scattered in the ice. These events are thus
characterized by hits spreading roughly spherically from the vertex, with a
preferred direction due to the Cherenkov cone.

We can exploit this topology of the signal hits. As demonstrated for
IceCube, requiring a local coincidence between photon hits is a very
efficient way to reduce the effect of random noise~\cite{icecube_LC}.
A hit is required to be accompanied by at least another hit within a
certain radius $r_\mathrm{RT}$ and time window $t_\mathrm{RT}$ in
order to fulfill the local coincidence criterion.
Fig.~\ref{fig:rtcleaning} shows an illustration of this radius-time
(RT) requirement.
%We have applied such a radius-time cleaning to the simulated events
%and find it to provide additional background rejection power on top
%of the phase space cut.
The hit cleaning was found to indeed reduce the noise trigger rate
while keeping most of the signal events. The parameters
$t_\mathrm{RT}$ and $r_\mathrm{RT}$ are optimized for the maximum
effective mass at each string spacing $d_\mathrm{str}$.

%In order to optimize the parameters $\Delta
%t$ and $r_\mathrm{RT}$, we apply the hit cleaning method to signal and noise events
%and then re-optimize the parameters of the phase space cut for the cleaned data
%set. The reduction of noise hits allows to increase $t_\mathrm{max}$
%for the same $f_\mathrm{noise}$ of $1\un{mHz}$.
%Fig.~\ref{fig:effmass_rtcomparison} shows the effective mass obtainable with
%this additional cleaning for $10\un{Hz}$ module noise to be increased by
%60\% from $7.5\un{Mton}$ to $12\un{Mton}$.

\subsubsection{Phase space cut (PS)\label{sec:phasespace}}
The local coincidence cleaning selects hits that are likely causally connected,
i.e.\ close to each other in time and space. However, this does not take into
account the global signal hit distribution in the detector. As photons
propagate away from the vertex, the photon distribution can be roughly modeled
as an expanding spherical shell. The rate of noise events can be significantly
reduced by defining a fiducial detector volume (the spherical shell) and
accepting only hits within this volume for the trigger.

\begin{figure}
    \includegraphics[width=\linewidth]{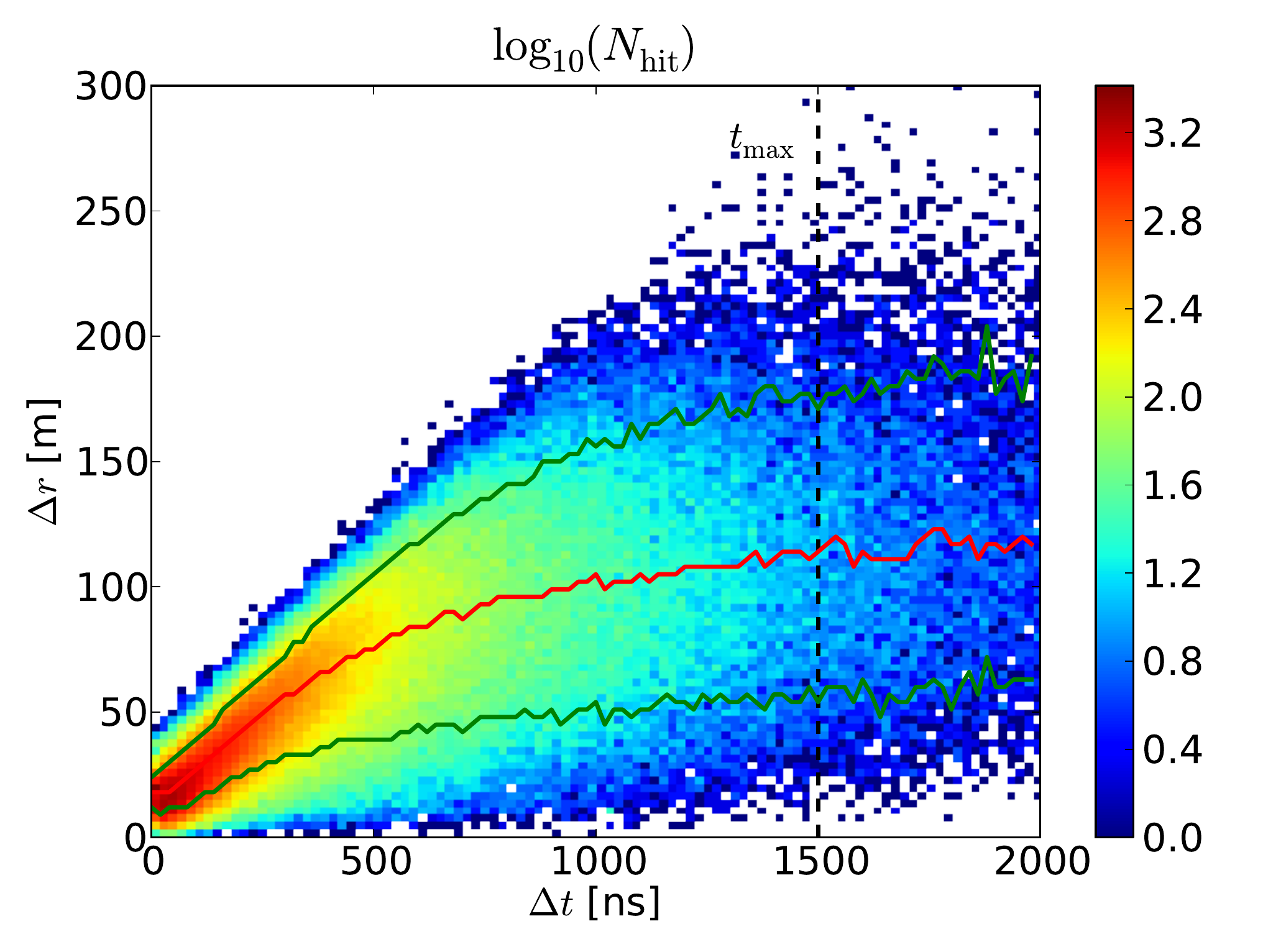}
    \caption
    {Distribution of signal hits in the simulated detector: spatial distance $\Delta r$
    to the vertex position vs.~time difference $\Delta t$ between the hit and the
    neutrino interaction. \label{fig:phasespace}}
\end{figure}

This fiducial volume is defined relative to the time and position of
the neutrino vertex. Since the vertex is not known a priori, we use a
$\chi^2$-minimization of the residual time, i.e.\ the photon
propagation time minus the expected photon propagation time for
straight travel from the vertex, to reconstruct the vertex position.
Using this simple method, a positional resolution of $\sim15\un{m}$
can be achieved. To incorporate this limited knowledge of the vertex
position, a random Gaussian smearing with $10\un{m}$ standard
deviation in each coordinate is applied to the vertex position of our
simulated events. Fig.~\ref{fig:phasespace} shows the distribution of
hits in the two-dimensional phase space (PS) given by the time between
the hit and the neutrino interaction and the distance between the hit
module and the reconstructed vertex. The 10\% and 90\% quantiles of
the distance are calculated for each time (green lines in
Fig.~\ref{fig:phasespace}, red line is the median) and give the inner
and outer radius of our fiducial volume
(cf.~Fig.~\ref{fig:rtcleaning}). The resulting spherical shell expands
with time and contains $\approx 80\%$ of the signal hits. Since all
photons are eventually absorbed, we cut the fiducial volume at time
$t_\mathrm{max}$.
%, which is a parameter optimized such that $f_\mathrm{noise}$ is limited to its required value (see Table~\ref{tab:fnoise}).\\

%\begin{figure}[tbp]
%    \begin{center}
%        \includegraphics[width=0.5\textwidth]{fnoise_vs_tmax_61strings_25m_50Hz}
%    \end{center}
%    \caption{Noise trigger rate $f_\mathrm{noise}$ of a detector with 61 strings and 25 m string
%    spacing in clear ice, for a module noise rate of $f_m=50\un{Hz}$, as function of phase space cut
%    time $t_\mathrm{max}$, for different values of $n_\mathrm{trig}$. Note the extremely steep rise
%    of the curve that spans 6 orders of magnitude in noise rate within only few hundreds of
%    nanoseconds.}
%    \label{fig:noiserate}
%\end{figure}

%We now reassess the noise trigger rate by limiting the hits that contribute to
%an event to those within this phase space region (cf.~Fig.~\ref{fig:rtcleaning}) for all times up to $t_\mathrm{max}$.
As the phase space volume (i.e.\ the number of sensors contributing to
the trigger) quickly increases with $t_\mathrm{max}$, the noise trigger rate
$f_\mathrm{noise}$ rapidly rises with $t_\mathrm{max}$ as well.
%Fig.~\ref{fig:noiserate} shows how steeply the noise
%trigger rate $f_\mathrm{noise}$ (calculated analytically) is rising with $t_\mathrm{max}$.
We choose the value of $t_\mathrm{max}$ so that it limits $f_\mathrm{noise}
\leq 1\un{mHz}$ for a given sensor noise rate $f_m$.

Under this constraint, we simultaneously optimize the number of hits
required to form a trigger $n_\mathrm{trig}$, the string spacing
$d_\mathrm{str}$ of the detector, the cleaning radius and time
($r_\mathrm{RT}$ and $t_\mathrm{RT}$), and $t_\mathrm{max}$ in order to maximize the
effective mass for SN signal neutrinos.
%All parameters are varied over
%sufficiently large ranges to ensure that the global maximum of the
%effective mass is found.

\begin{table}[b!]
  \begin{center}
    \begin{tabular}{rlcccc}
        \toprule
         & $f_m$                           & [Hz]    & 10    & 50    & 500  \\ \midrule
        \multirow{5}{*}{\rotatebox{90}{\parbox[t]{2cm}{\centering
        optimized parameters}}}
         & $d_\mathrm{str}$                & [m]     & 25    & 25    & 25 \\
         & $n_\mathrm{trig}$               &         & 5     & 5     & 5 \\
         & $r_\mathrm{RT}$                 & [m]     & 115   & 85    & 40 \\
         & $t_\mathrm{RT}$                 & [ns]    & 1200  & 400   & 250 \\
         & $t_\mathrm{max}$                & [ns]    & 1590  & 850   & 390 \\ \midrule
        \multirow{3}{*}{\rotatebox{90}{\parbox[t]{1.2cm}{\centering
        prop\-erties}}}
         & $f_\mathrm{noise}$              & [Hz]    & 1.0   & 3.4   & $2.2\cdot10^4$ \\
         & $f_\mathrm{noise}^\mathrm{cut}$ & [mHz]   & 1     & 1     & 1 \\
         & $M_\mathrm{eff}$                & [Mton]  & 13.1  & 10.5  & 5.4 \\
        \bottomrule
    \end{tabular}
    \caption{Optimal parameters for string spacing ($d_\mathrm{str}$), required
    hit multiplicity ($n_\mathrm{trig}$), RT cleaning radius ($r_\mathrm{RT}$)
    and time ($t_\mathrm{RT}$) as well as maximum hit time window
    ($t_\mathrm{max}$) together with resulting effective mass $M_\mathrm{eff}$
    as function of module noise rate $f_m$. The parameters are chosen such that
    RT cleaning and phase space cut limit the rate of noise triggers
    $f_\mathrm{noise}$ down to $f_\mathrm{noise}^\mathrm{cut} = 1\un{mHz}$.
    \label{tab:pscut}}
  \end{center}
\end{table}

\subsubsection{Conclusion}
\begin{figure}
\includegraphics[width=\linewidth]{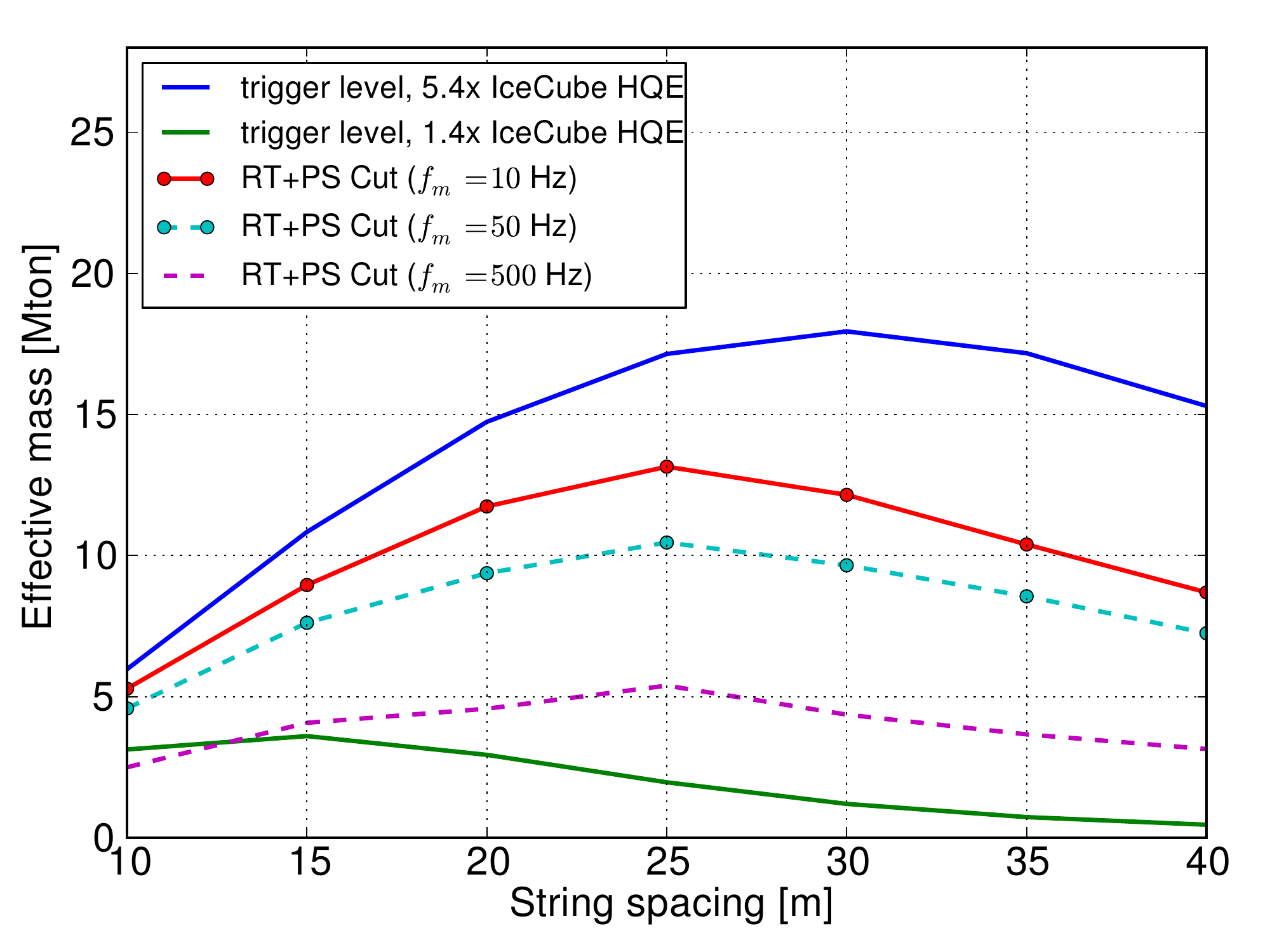}
\caption
   {Effective mass at trigger level with modules having 1.4 (lower solid) and
   5.4 times the effective area of IceCube HQE DOMs (upper solid). The same
   after applying radius-time (RT) cleaning and phase space (PS) cut for
   different module noise rates (solid with bullets, dashed). The noise
   cleaning is done such that a noise trigger rate of about $\unit[1]{mHz}$
   remains, a dead time of 0.16\% due to atmospheric muons is included.
   \label{fig:effmass_pscut}}
\end{figure}

%We simulate the detector in the clear ice as described above with 127 strings of
%300\un{m} length (c.f.~Fig.~\ref{fig:geo}).
Fig.~\ref{fig:effmass_pscut} shows the resulting effective mass after
the two self-noise cuts have been applied, Tab.~\ref{tab:pscut}
lists the optimal parameters as function of module noise $f_m$. While
this method reduces the trigger efficiency for neutrino interaction by
up to a factor of about 4, at the same time the noise trigger rate is
reduced by many orders of magnitude.
For the module dark noise rate of $500\un{Hz}$, as provided by IceCube
DOMs, the effective mass of this detector configuration falls short of
the initial target of $10\un{Mton}$. In order to retain $10\un{Mton}$,
reducing the sensor self-noise rate to $\lesssim50\un{Hz}$ per meter
instrumented string ($350\un{mHz/cm^2}$ effective photo-sensitive area) is desirable.

%Equally important as a low noise rate is the effective area of the optical
%modules: Even with modules 36\% larger than IceCube high
%quantum efficiency (HQE) DOMs deployed each meter of a string, less
%than $\unit[5]{Mton}$ are achieved at trigger level, before cutting
%away the noise events (low solid line in Fig.~\ref{fig:effmass_pscut}).

%Operation of a very dense detector at the lowest possible trigger threshold
%introduces a strong dependence of the trigger rate on the module noise.
%Algorithms exploiting the spatial and temporal distribution of the light
%%(that
%%cannot be used in the diffusive ice)
%have to be employed to obtain a sufficiently
%low rate of false SN detections.
While more advanced algorithms considering the full event topology would still
allow for a moderate increase of the effective mass, this study demonstrates
the importance of both large photo-sensitive area and at the same time low noise
optical sensors, posing quite a technological challenge.
Correlated noise that was not treated here will be more difficult to reject and
provides additional motivation to seek low dark noise rates in future
photo-sensor R\&D. In the following, we assume a module noise rate of $\unit[10]{Hz}$.

% Produced by: reco/PhaseSpaceModular_plot.py
%\begin{figure}
%\includegraphics[width=\linewidth]{effectiveMass_vs_spacing_127strings_afterPSCut_smearing}
%\caption
%  {Effective mass of the simulated 127-string detector at trigger level (solid)
%  and after a phase-space cut is applied to limit the false trigger rate to
%  1\un{mHz} for module noise rates of $500\un{Hz}$ (dashed), $50\un{Hz}$
%  (dashed, circles) and $10\un{Hz}$ (solid, circles).\label{fig:effmass_pscut}}
%\end{figure}

% Produced by: reco/PhaseSpaceModular_plot.py

\subsection{Muon background}
\label{sec:muon-back}
Muons crossing the detector or passing nearby are easily separable from the SN
neutrino signal via the huge amount of Cherenkov light produced by the extended
track. An additional outer layer of photosensitive modules, naturally provided
by IceCube, will ensure that even muons passing by at large distances or
stopping just above the detector can be identified as such and be vetoed.

For a conservative first estimate of the dead time caused by atmospheric muons, we assume all
muons reaching the top of the detector being energetic enough to cross it.
%%% For deep ice:
From~\cite{BugaevEtAl} we obtain a muon
flux of $\Phi_\mu = 8 \cdot 10^{-8}\un{cm^{-2} s^{-1} sr^{-1}}$ for the top of the
% clear ice
detector at 2150 m depth, giving a muon passing rate of about
\begin{equation}
  R_\mu = \Phi_\mu \cdot 300\un{m} \times 300\un{m} \cdot \pi \approx 230\un{Hz}
\end{equation}
using a detector cross-sectional area of $(300\un{m})^2$ and an
effective solid angle of $\pi$ that accounts for the lower flux from
angles closer to horizon where the ice shield is thicker. A muon
traveling through the entire detector has a track length of $\approx
300\un{m}$ and emits about $N_0 \approx 10^7$ Cherenkov photons on its
path ($\approx 360$ photons per cm). We assume as a worst case that
all photons are trapped within the detector volume by scattering to
compute how long the photons will remain detectable within the
detector before they are absorbed. The number of photons after a path $x
= c\,t/n$ is given by:
\begin{equation}
  N_\gamma = N_0 \cdot e^{-\frac{t}{t_\mathrm{abs}}},
\end{equation}
with $t_\mathrm{abs} = n \, \lambda_{a}/c = 0.44 \cdot 10^{-6}\un{s}$, for an absorption length of
$\lambda_{a} \lesssim 100\un{m}$
%(as found in the clear ice)
and a refractive index of $n=1.31$. After a time
\begin{equation}
  t = t_\mathrm{abs} \ln\frac{N_0}{N_\gamma} \approx 7\un{\mu s}
\end{equation}
the number of unabsorbed photons in the detector due to a crossing muon has fallen below $N_\gamma = 1$
which is well below our trigger threshold. We therefore
conservatively assume each passing muon to illuminate the detector for a time
interval of $7\un{\mu s}$ which can be considered as
dead time for supernova neutrino detection, corresponding to a fraction of
$R_\mu \cdot 7\un{\mu s} = 0.16\%$ of detector operation time.

Passing muons are easily identified because of the dense
instrumentation and can be rejected applying a veto. However, muons
also produce spallation products via fragmentation of $^{16}$O nuclei
and capture of the generated neutrons~\cite{Bays:2011si}. The decay of
these numerous products can mimic low-energy neutrino events and is a
serious background. Super-Kamiokande has
demonstrated~\cite{Bays:2011si, Malek2003} that a cut on a likelihood
function including spatial and temporal distance from the passing muon
as well as energy loss of the muon can be used to efficiently remove
spallation events. However this results in an additional $20\%$
effective dead time for Super-Kamiokande while raising the threshold
for neutrino detection to $\sim 15\un{MeV}$. As not all of the
spallation and neutron capture cross-sections are known and their
calculation goes beyond the scope of this work, we cannot quantify
this background. Yet, we note that Super-Kamiokande is located at a
depth of $2700$ meters water equivalent~\cite{SuperK-300days},
comparable to the deep location considered here, and the passing muons
should be equally well reconstructed, indicating that performance
factors may be similar. A possible demonstration that this background can
be controlled is left for future studies.
%%%%% old for deep and shallow side-by-side %%%%%
%For the shallow
%diffusive ice around 900 meters depth in contrast, where the flux of muons is
%larger by a factor of twenty, suppression of the spallation backgrounds may prove
%signifcantly more challenging.

\subsection{Solar neutrino background}
\label{sec:solar}
Put aside the cosmic neutrino background---which is too low in energy
to be detectable---the by far dominant flux of neutrinos at Earth
comes from the Sun, where neutrinos are abundantly produced in several
different fusion cycles~\cite{GiuntiKim}.  As only $\nu_e$ and no
$\bar\nu_e$ are generated in the Sun, solar neutrinos cannot undergo
inverse beta-decay (IBD).  The dominant interaction for solar neutrinos is
elastic scattering on electrons (ES). Charged current interactions on
oxygen nuclei are about two orders of magnitude less
important~\cite{Haxton1999} and are ignored here.  Furthermore, only
$^8$B neutrinos need to be considered, since all other solar neutrino
fluxes are too low in energy to be detectable in our configuration or
well below the $^8$B flux in magnitude~\cite{Bahcall2004,Bahcall2005}.

\begin{figure*}
    \begin{center}
        \subfloat[]{\includegraphics[width=0.45\textwidth]{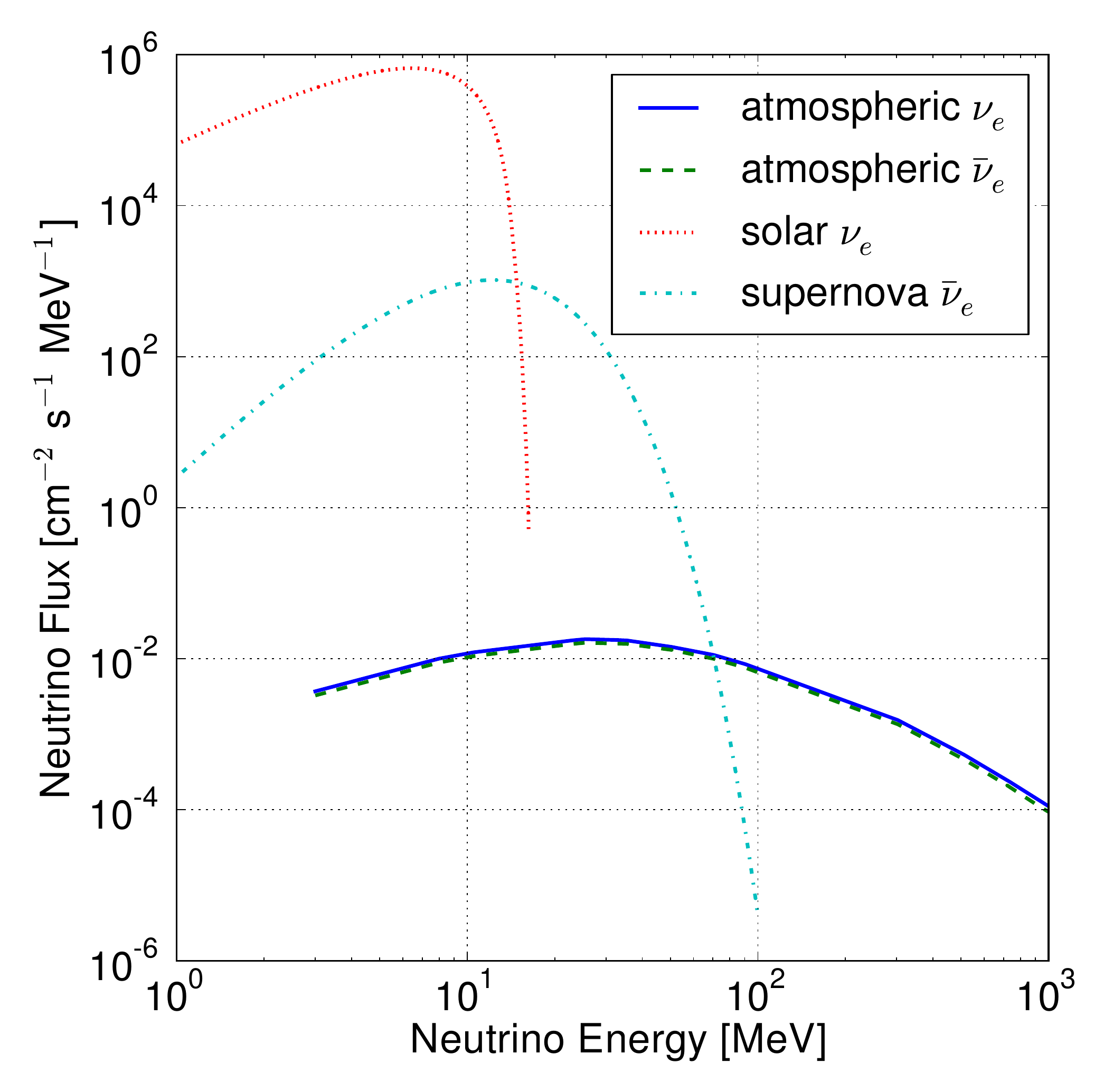}}
        \subfloat[]{\includegraphics[width=0.45\textwidth]{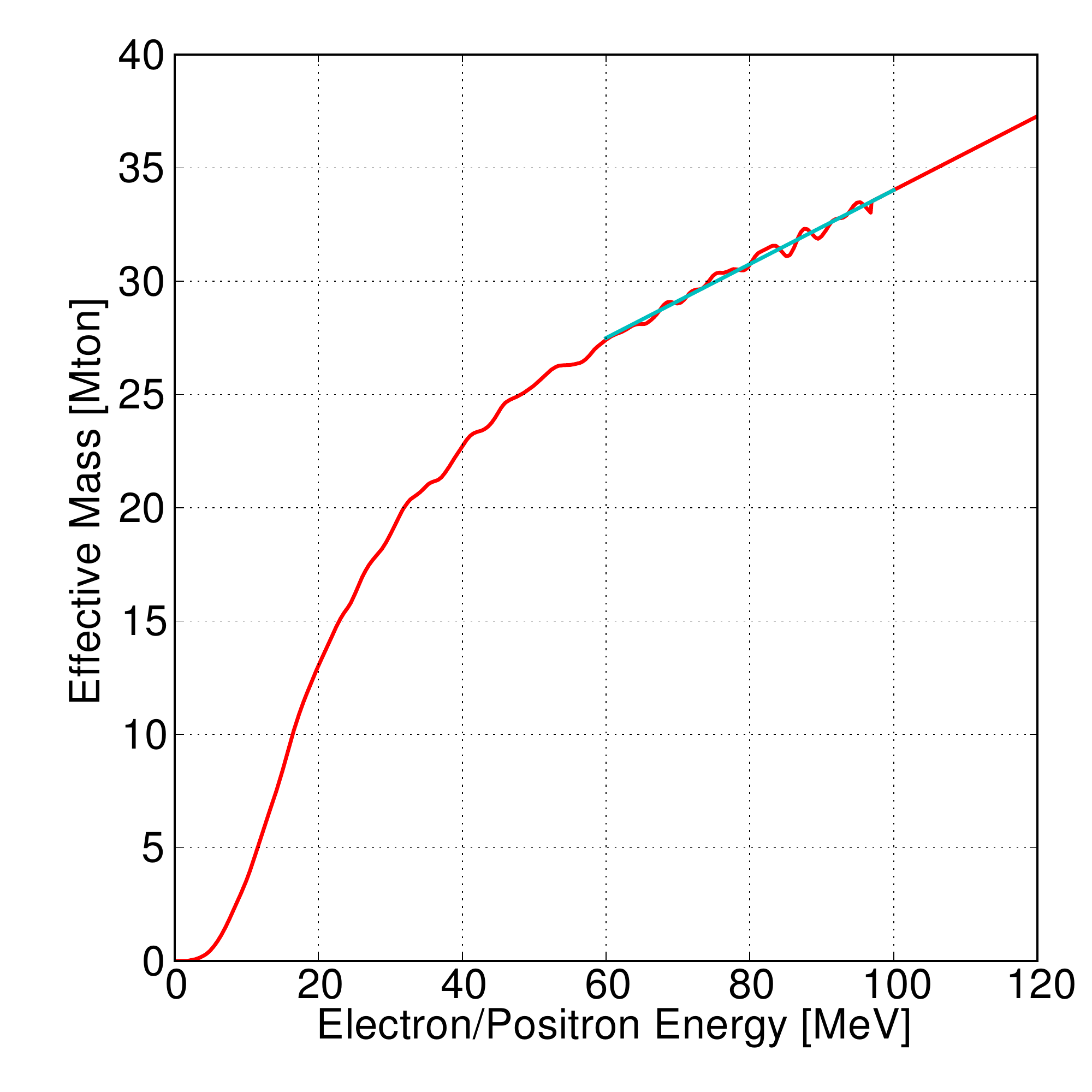}} \\
        \subfloat[]{\includegraphics[width=0.45\textwidth]{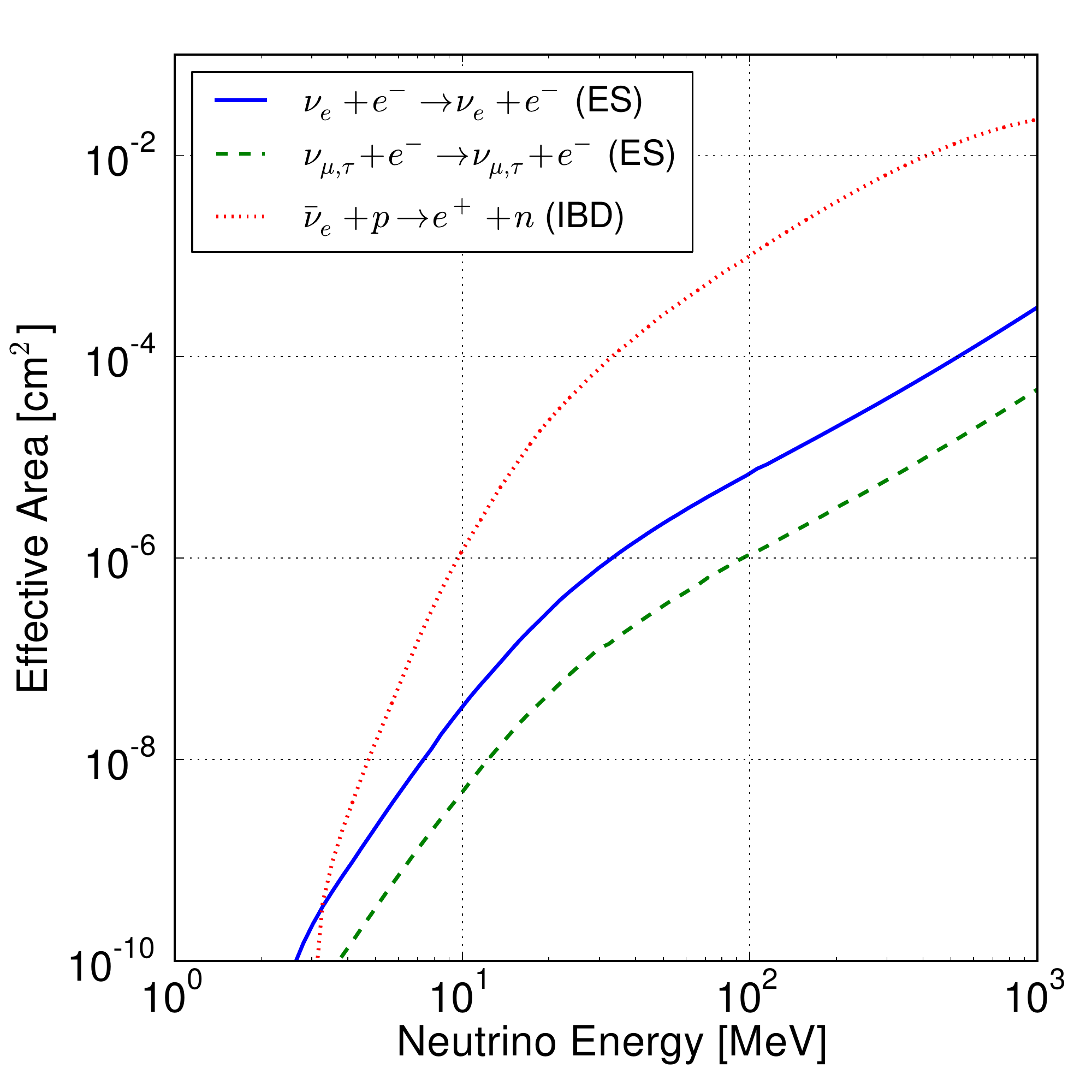}}
        \subfloat[]{\includegraphics[width=0.45\textwidth]{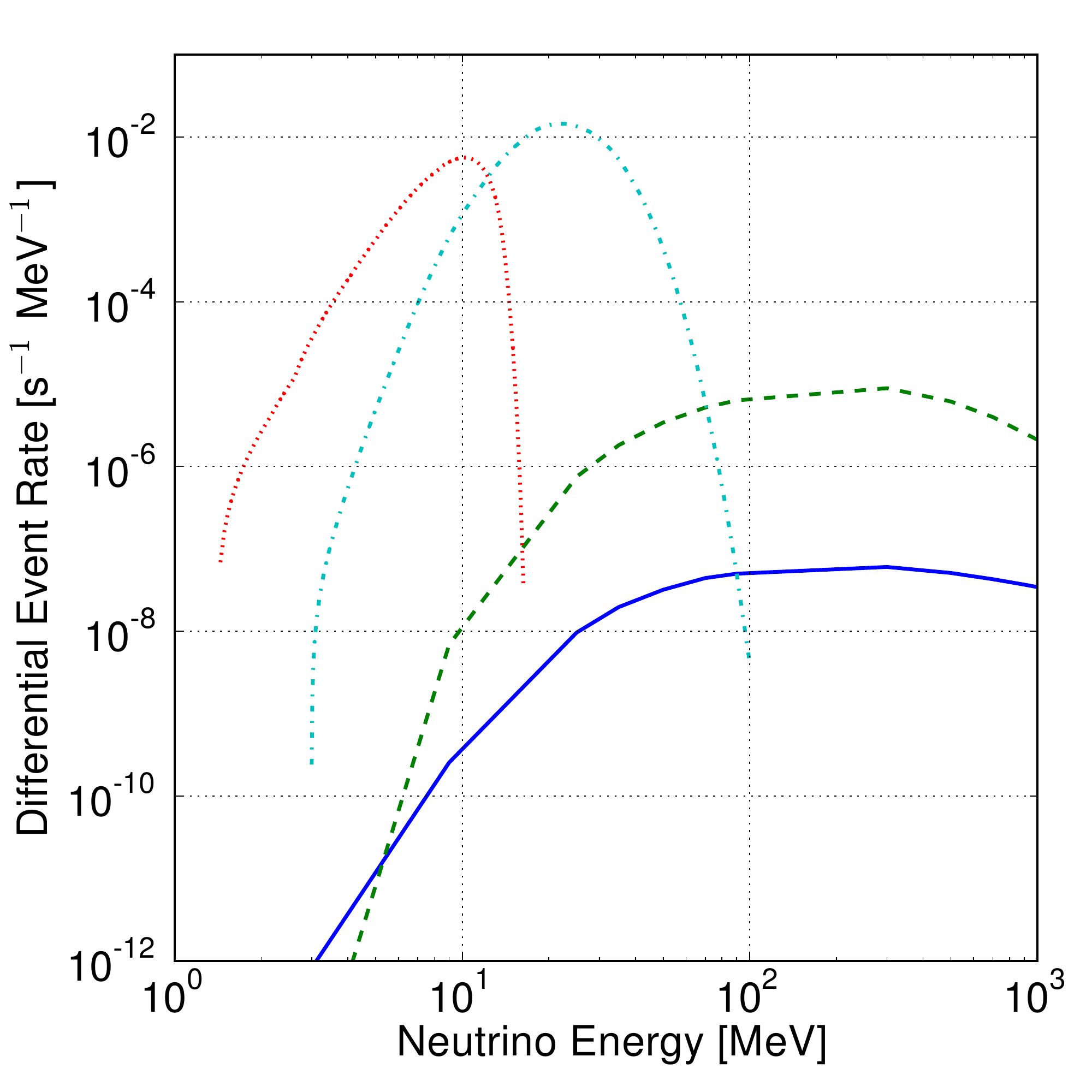}}
    \end{center}
    \caption{(a)~The flux for solar~\cite{Bahcall1996,Haxton1999} and
    atmospheric~\cite{Gaisser1988} neutrinos as function of
    energy. The Supernova neutrino flux on Earth according
    to the LL model~\cite{livermore}, normalized to an average power of $4.9 \cdot 10^{51}
    \un{erg\,s^{-1}}$ in $\bar\nu_e$ for a supernova distance of $10\un{Mpc}$
    is shown assuming a burst duration of 10\un{s}.
    (b)~Effective mass as function of $e^{\pm}$ energy,
    triggering on 5 hit modules, after applying noise cleaning down to
    $f_\mathrm{noise}^\mathrm{cut}=\unit[1]{mHz}$ for $\unit[10]{Hz}$ module noise.
    Above $100\un{MeV}$, a linear extrapolation (dashed) is used.
    (c)~The effective area of the same detector for elastic scattering (ES)
    and inverse beta-decay (IBD), not including cuts to reject solar neutrinos.
    (d)~Event rate per $\un{MeV}$ as function of neutrino energy for
    solar, atmospheric and supernova neutrino fluxes from (a) with
    effective area from (c).
    }
    \label{fig:backgrounds}
\end{figure*}

To calculate the interaction rate, we use an analytical expression for neutrino-electron elastic scattering (ES) from Eqn.~(5.25)
in~\cite{GiuntiKim}
\begin{equation}
    {\textstyle \frac{d \sigma}{d T_e}(E_\nu,T_e) = \frac{\sigma_0}{m_e} \left[ g_1^2 + g_2^2
    \left( 1-\frac{T_e}{E_\nu} \right)^2 - g_1 \, g_2 \, \frac{m_e\,T_e}{E_\nu^2} \right] }
    \label{eqn:es}
\end{equation}
with the kinetic energy of the recoil electron in the laboratory frame, $T_e$, and
\begin{align*}
    \sigma_0 &= \frac{2 \, G_F^2 \, m_e^2}{\pi} \simeq 88.06 \cdot 10^{-46} \un{cm^2} \\
    g_1 &=
    \begin{cases}
        \frac{1}{2} + \sin^2\theta_W \simeq 0.73 \quad \mbox{for $\nu_e$} \\
        -\frac{1}{2} + \sin^2\theta_W \simeq -0.27 \quad \mbox{for $\nu_{\mu,\tau}$}
    \end{cases} \\
    g_2 &= \sin^2\theta_W \simeq 0.23
\end{align*}
and fold it with the energy-dependent effective mass of our detector given in
Fig.~\ref{fig:backgrounds}(b). Taking the shape of the $^8$B neutrino spectrum
from~\cite{Bahcall1996} and normalizing it to a total flux of
$\Phi(^8\mathrm{B}) = 5 \cdot 10^6 \un{cm^{-2} s^{-1}}$~\cite{SNO_total8BFlux}
with components $\Phi_{\nu_e}(^8\mathrm{B}) = 1.7 \cdot 10^6 \un{cm^{-2}
s^{-1}}$ and $\Phi_{\nu_{\mu,\tau}}(^8\mathrm{B}) = 3.3 \cdot 10^6 \un{cm^{-2}
s^{-1}}$~\cite{SNO_eAndMuTau8BFlux}, we arrive at an approximate solar neutrino
event rate of $65\un{mHz}$.
%for a detector in the clear ice at trigger level.
Additional application of local coincidence cleaning
(cf.~Sec.\ref{sec:rtcleaning}) and the phase space cut
(cf.~Sec.\ref{sec:phasespace}) reduces this rate to $30\un{mHz}$, still
significantly larger than the maximum allowed rate of random background events
$f_\mathrm{noise}^\mathrm{BG} = 1\un{mHz}$.

However, a number of methods can be applied to further diminish the rate of
solar neutrino events. The bulk of solar neutrinos is less energetic than the
bulk of SN neutrinos (cf. Fig.~\ref{fig:backgrounds}\,(a)). Changing the
trigger requirement from 5 hits to 7 hits increases the energy
threshold and thus reduces the expected event rate of solar neutrinos by a
factor of three, while reducing the signal efficiency for SN neutrinos
from the LL model by only $30\%$.

Alternatively, the electron emerging from the
ES roughly keeps the direction of the incident neutrino, while the
IBD effectively randomizes the positron direction. For a
sufficiently densly instrumented array such as Super-Kamiokande, an angular
resolution of about $\pm30^\circ$ at $E_e=10\un{MeV}$ is feasible~\cite{Haxton1999,
Malek2003}. Assuming a one-sigma (68\% of the events) angular cone of
this size to reject neutrinos from the direction of the Sun, we cut
away a solid angle of $\Omega_\text{cut} = 2 \pi (1 - \cos30^\circ)$
of the sky. We use this (instead of a cut on number of hits) to discriminate the
solar neutrino rate to $f_\odot = (1-0.68)^3 \cdot 30\un{mHz} \approx
1\un{mHz}$, while retaining
%%% $(1- (60^\circ)^2 / \Omega_{sky})^3 \gtrsim 93\%$
% in python: ( 1 - (pi*30**2 / 40000.) )**3 = 80 % ???
% more accurate:
% Raumwinkel, der ausgeschnitten wird: Kegel mit 30 Grad Öffnungswinkel
% Omega_cut = 2 \pi (1 - \cos30\deg)
% Anteil von IMB-Events (random direction), die wir behalten:
% 1 - \Omega_cut / (4\pi) = 1 - 2 \pi (1 - \cos30\deg) / (4\pi) = 0.93301270189221941
% SN-Events, die wir behalten: ( 1 - \Omega_cut / (4\pi) )^3 \approx 81 %
$( 1 - \Omega_\text{cut} / (4\pi) )^3 \approx 81\%$
of the SN events. While more efficient than a cut on the number of hits, the
most powerful rejection will be achieved using a likelihood method that
incorporates both the direction and energy for each event
within the time window. In absence of a full reconstruction we refrain from a
more detailed discussion, but note it as a requirement to reduce the large rate of solar neutrinos.
%%%%% old for deep and shallow side-by-side %%%%%
%It is also worth mentioning that a fundamental
%prerequisite is the ability to reconstruct the direction of the emerging lepton.
%, which
%can only be done in the clear ice where most photons are only slightly deflected
%before absorption. In the diffusive ice, all directional information is lost
%during the random walk of the photons.

% backgrounds/estimateBackgroundRates.py -f sim/effectiveMassHistograms/EffectiveMass-25m-127String-PSRTCut_5Hits.root --fit --smooth --effarea

\begin{table}[tbp]
  \begin{center}
    \begin{tabular}{cccc}
      \toprule
      {\bf events per $\mathbf{10\un{s}}$} & {\bf solar $\mathbf{\nu_e}$} &
      {\bf atm. $\mathbf{\nu_e + \bar\nu_e}$} & {\bf LL $\mathbf{\bar\nu_e}$ }\\
      \midrule
      %5 hits, no cleaning & 0.718 & 0.0038 & 3.739 \\
      %5 hits, RT+PS       & 0.337 & 0.0032 & 2.755 \\
      %6 hits, RT+PS       & 0.202 & 0.0031 & 2.335 \\
      %7 hits, RT+PS       & 0.109 & 0.0029 & 1.909 \\
      5 hits, no cleaning & 0.65  & 0.004  & 3.74  \\
      5 hits, RT+PS       & 0.30  & 0.003  & 2.76  \\
      6 hits, RT+PS       & 0.18  & 0.003  & 2.34  \\
      7 hits, RT+PS       & 0.10  & 0.003  & 1.91  \\
      \bottomrule
    \end{tabular}
    \caption[Background rates of simulated detectors]{Average number of events
    per 10\un{s} from solar and atmospheric neutrino background (up to 100 MeV)
    as well as the SN neutrino signal for a LL supernova in $10\un{Mpc}$ distance
    for different numbers of hit modules and with and without noise cleaning (RT+PS) applied.
    %The LL flux
    %was integrated from $0$ to $100\un{MeV}$, the atmospheric fluxes from $3$ to $100\un{MeV}$ and the
    %solar flux from $0.02$ to $16.34\un{MeV}$.
    % For these numbers, an integration up to $100\un{MeV}$ was done. %%% Not anymore!!! (atmo up to 2500 MeV)
    }
    \label{tab:backgrounds}
  \end{center}
\end{table}

\subsection{Atmospheric neutrino background}
Cosmic rays colliding with the Earth's atmosphere produce $\nu_e$ and $\bar\nu_e$ in similar
abundance.
%, and a larger number of $\nu_\mu$ and $\bar\nu_\mu$.
%which cannot be seen in our detector at energies of few tens of MeV
%because they are too low-energetic to produce a muon.\pdfmargincomment[avatar=MV,id=42]{But
%what about NC interaction of NuMu?}
%\pdfreply[avatar=LS,replyto=42]{Da ist der WQ vermutlich immer noch zu klein, als dass das relevant w\"{a}re}
%High energy muon neutrinos produce muons that also
%need to be vetoed, adding a tiny fraction to the dead time mentioned in section
%\ref{sec:muon-back}.
The dominant component are the electron anti-neutrinos, interacting via IBD
with a cross-section two orders of magnitude higher than the ES of electron
neutrinos~\cite{Malek2003}. Taking the atmospheric neutrino flux calculations
from~\cite{Gaisser1988}, the ES cross-section from Eqn.~\ref{eqn:es} and the
cross-section for IBD given in the phenomenological parametrization
in~\cite{strumia} (``Na\"ive +'' model), and integrating the event rate from
$3-100\un{MeV}$, we arrive at an expected trigger level event rate of
$0.004\un{mHz}$ for $\nu_e$ and $0.4\un{mHz}$ for $\bar\nu_e$ triggering on 5
hit sensors.
%%% XXX: forgot the contribution of elastic scattering to the solar \bar\nu_e event rate?
%in the clear ice.
The resulting spectrum, also shown in
Figure~\ref{fig:backgrounds}\,(d), peaks well above the peak of SN interactions,
allowing to further discriminate these events. While somewhat more abundant,
ES of $\nu_\mu$ on $e^-$ is only possible via neutral current
interactions, and thus has a factor six smaller cross-section.

%%%%%%%%%%%%%%%%%%%%%%%%%%%%%%%%%%%%%%%%%%
% Talk about atm. numu? > 50 MeV: visible, decay to electrons -> 'sub-events'
%                                    sub-event cut and Ch. angle cut
%                       < 50 MeV: invisible muons dacaying to Michel spectrum electrons
%%%%%%%%%%%%%%%%%%%%%%%%%%%%%%%%%%%%%%%%%%

Another component of the background are \emph{invisible muons} that are
produced in the interactions of low-energy atmospheric muon neutrinos. The
muons themselves are below the threshold for Cherenkov light emission and thus
invisible. They can only travel few tens of centimeters and then decay to electrons
which---due to their lower mass---can have a velocity above the Cherenkov
threshold and become visible. These \emph{Michel electrons} have been measured
by Super-Kamiokande~\cite{Beacom2010, Malek2003} and amount to $\sim 90$ events
per year in their effective volume of 22.5\un{ktons}. At the peak energy of the
Michel electron spectrum, $\unit[40]{MeV}$, the effective mass of the detector
simulated in this work is about 1000 times larger (see
Fig.~\ref{fig:backgrounds}(b), so we can expect about $90000$ events per year,
or a rate of $\approx 3\un{mHz}$. While already small compared to the
background of solar neutrinos, further reduction can be achieved by using the
surrounding IceCube detector to veto accompanying atmospheric muons. Note that
only invisible muons from muon neutrinos and not the atmospheric muons
themselves can penetrate the detector and produce Michel electrons as, once below the
Cherenkov threshold, the muons will decay within $\sim\unit[1]{m}$.
%in the deep ice.
% reference to 'can be used for energy calibration'
%~\cite{SuperK-solarspec} energy calibration sources:
%   1. muon dacay electrons   2. spallation products   3. 16N producing by stoppong muon capture on oxygen

% \pdfcomment[avatar=MV]{In arXiv:1111.5031, they say that Michel electrons only come from numu
% CC reactions of atmospheric neutrinos. Why are Michel electrons not dominantly from the atmospheric muons
% that go below Cherenkov threshold close above the detector?}
% -> Nachdem ein Myon unter der Cherenkov-Schwelle ist, kommt es noch ~1 m weit...

\subsection{Summary}
The large sensor multiplicity requires intelligent trigger and selection
algorithms to cope with the backgrounds arising from sensor self-noise,
atmospheric muons, solar as well as atmospheric neutrinos.
%In this challenging background situation, a detector in the diffusive ice close to the surface is
%strongly disfavored due to a) the inability to reject self-noise triggers by
%their hit distribution pattern, b) the much higher muon background
%leading to significant downtime, c) the inability to use the surrounding IceCube and DeepCore detectors as muon veto
%and d) the solar neutrino backgrounds which cannot be vetoed
%effectively due to the lack of directional reconstruction.
Despite making use of the event topology, the suppression of
self-noise to a sufficiently low level will be a major challenge and
requires future improvements in sensor development. Vetoing of
atmospheric neutrinos and muons will result in some downtime in the detector.
In particular, the discrimination of spallation products from muons passing the
ice may pose a significant challenge and still has to be demonstrated. The
dominant source of {\it physical} \/(i.e.\ neutrino) background stems from solar
$^8\mathrm{B}$ neutrinos, that need to be suppressed by reconstructing their
direction and/or increasing the energy threshold.

%==============================================================================
\section{Alternative detector location}
\label{sec:diffusive}
%==============================================================================

An interesting alternative location for a Cherenkov detector in the
South Pole ice is at a depth of $750-1050\un{m}$. It is known from
measurements that at this depth the absorption length is exceptionally
large with up to $\lambda_a \approx
350\un{m}$~\cite{porrata}. However, the
presence of air bubbles results in a very short scattering length of only $\lambda_e \approx
0.3\un{m}$~\cite{porrata}. This results in an effective propagation
length of $\lambda_p = \sqrt{\frac{\lambda_e \cdot \lambda_a}{3}} \sim
6\un{m}$ after which the photon flux has dropped by a factor
$e^{-1}$~\cite{absorption}. The photons cannot travel large
distances and are hence confined to a small volume for a rather long
time of $\lambda_a/c \approx 1 \un{\mu s}$ before finally being
absorbed. This leads to a large detection probability, given the light
is emitted in the vicinity of a photo-sensor. The achievable
effective mass at trigger level is thus several times
larger than in the deep ice.

We have simulated such a detector (same detector layout as in the deep ice)
with an analytic description of the photon propagation (a random walk) using
perfectly efficient cylindrical modules with $1\un{cm}$ circumference and
$1\un{m}$ length, having an effective area comparable to HQE IceCube DOMs,
roughly a factor
% 4 times % (factor 4 difference would mean factor 1.84 between cylinder and IceCube DOM)
% (this seems to be wrong, since cylinder has projected area of 25 cm^2 and DOM has 19.4 cm^2,
% thus, the factor is rather ~1.3)
5.4 smaller
% real number: ~142/25 ~= 5.7
than what was used in the deep ice. After applying the same optimized
module noise cuts to reduce the self-noise rate to $\unit[1]{mHz}$,
one still obtains $\sim\unit[14]{Mton}$ effective mass, when
triggering on 5 hit modules (including the higher dead time induced by
atmospheric muons, conservatively estimated to 14\% as explained
in~\ref{sec:muon-back}).

The main drawback of the shallow, diffusive ice is the lack of
directional reconstruction of events. Due to the strong scattering,
any information on the direction of the charged particle is lost.
Among other things, this makes it impossible to veto elastic
scattering events from solar neutrinos by their direction. Thus, a cut
on the number of detected photons---which is directly proportional to
the event energy and can be used as an energy proxy---is the
only option to suppress that background. But due to the large overlap
of the energy spectra of solar neutrinos and SN neutrinos (see
Figure~\ref{fig:backgrounds}) one significantly loses detection
efficiency.

\begin{figure}
\includegraphics[width=\linewidth]{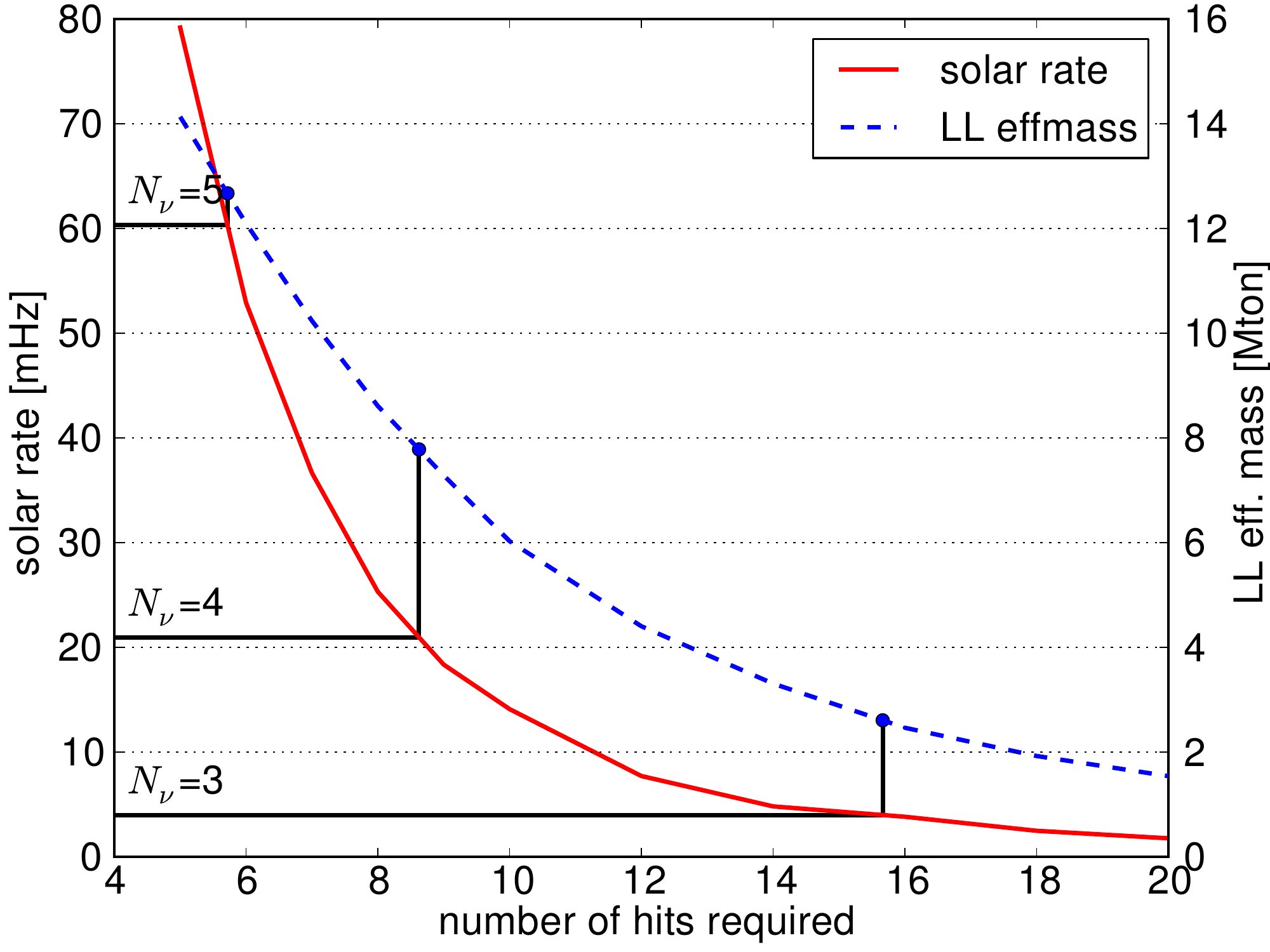}
% this plot is for a detector with 30 m spacing (optimum), 10 Hz
% module noise, no RT cleaning was done, but the phase space cut. RT
% not needed in diffusive ice at 10 Hz, only for higher module noise
\caption
   {The solar neutrino event rate in the shallow ice detector (from
   neutrino-electron elastic scattering), together with the effective
   mass for SN neutrino events (Lawrence-Livermore), as a function of
   the number of hit modules required to trigger an event (which is a
   proxy of event energy). Horizontal lines indicate tolerable noise
   rates for a 1 second time window during which at least $N_\nu$
   events shall make up a SN detection. The vertical lines point to
   the corresponding resulting effective mass.
   \label{fig:shallow-solar}}
\end{figure}

Figure~\ref{fig:shallow-solar} shows the solar neutrino event rate and
the effective mass for SN neutrinos (LL spectrum) as function of the
number of hit modules required for the event trigger. Horizontal lines
show the levels of solar event rate that can at most be tolerated,
with the SN search window reduced from 10 seconds to 1 second (thus
missing about 40\% of the SN neutrino events, according to
Figure~\ref{fig:nuflux}). For a number of neutrino events $N_\nu \geq
3$ per SN detection, a solar rate of $\approx\unit[4]{mHz}$ can be
allowed (see Eqn.~\ref{eqn:fnoise}). One would have to select events
with at least 16 hit modules and the effective mass drops to about
$\unit[2.5]{Mtons}$. Increasing the number of instrumented modules or
the modules' photo-effective area will not suffice to recover the
effective mass, because the solar event rate would be increased as
well.

Raising the neutrino event threshold $N_\nu$ relaxes the requirement
on the solar rate---i.e.\ the cut on the hit modules---yielding a
higher effective mass, but at the same time reducing the number of
detected SNe. Optimizing again over all parameters, we find the best
results for the shallow ice at $N_\nu = 5$ with a cut on at least 6
hit modules (c.f.~Figure~\ref{fig:shallow-solar}) in a SN search
window of $\Delta t = \unit[1]{s}$ at an effective mass of about
$\unit[12]{Mton}$. However, only 60\% of the SN neutrino events arrive
within $\Delta t = \unit[1]{s}$, (c.f.~Figure~\ref{fig:nuflux}),
adding a factor $0.6^5 \approx 8\%$ to the supernova detection rate.
In contrast, in the deep ice we are already at the optimum for a
minimal $N_\nu$ of 3, with a SN search window of $\Delta t =
\unit[10]{s}$ with a collection efficiency of nearly 100\%, yielding
an effective mass of $\unit[13.1]{Mton}$. The achieved SN detection
rates are compared in Table~\ref{tab:detections}, revealing that the
shallow ice is inferior to the deep ice.

%It shows the SN detection
%rate as function of the $N_\nu$ threshold for the optimal detector in
%the deep clear ice and for the optimal detector in the shallow
%diffusive ice. Self-noise reduction cuts were applied (to a level of
%$\unit[1]{mHz}$). For the deep ice, a directional cut for removal of
%solar events with efficiency of 81\% was assumed (see
%Section~\ref{sec:solar}). For shallow ice, a cut of
%$N_\mathrm{hit}\geq16$ was applied in order to reduce the solar rate
%to $\approx \unit[4]{mHz}$, which is tolerable searching for
%$N_\nu\geq3$ events within a search window of 1 second (efficiency of
%$\approx 60\%$, c.f.~Figure~\ref{fig:nuflux}, taken into account).
%Going from $N_\nu\geq3$ to $N_\nu\geq4$ events results in a loss of about
%36\% of all SN detections. At the same time, one can relax the
%$N_\mathrm{hit}$ cut to 9 hit modules, increasing the effective mass
%by about a factor 3 (to $\unit[8.5]{Mton}$). Going even further to
%$N_\nu\geq5$, one can allow events with 6 hit modules and has a
%roughly 5 times higher effective mass than at $N_\nu\geq3$, however
%reducing the SN detections to 45\% of the $N_\nu\geq3$ detections.

Other physics cases such as proton decay also rely on directional information
and cannot be pursued in the shallow ice. Additionally, the detector in the
shallow ice will suffer more from other backgrounds as well, above all the
atmospheric muons (much longer dead time) and muon-induced spallation events
that might even become unmanageable in the low depth. Also, IceCube cannot be
used as a veto in the shallow ice. Therefore, we disfavor a detector located
in the shallow, diffusive ice.

%In the diffusive ice the light propagation is well described analytically as a
%random walk with absorption~\cite{absorption}, allowing for convenient
%simulation of the photon tracking.

%\begin{figure}
%\includegraphics[width=\linewidth]{SNDetectionRate_vs_nu_event_includingDifferentEffMassesOfNuEventThresholds}
%\caption
%   {Detection rate of SNe with $N_\nu$ or more events as function of
%   $N_\nu$, for deep and shallow ice (in deep ice not continued beyond
%   $N_\nu=5$). The effect of increasing effective mass due to relaxed
%   noise cuts (going to higher $N_\nu$) is included.
%   \label{fig:sn-detections}}
%\end{figure}

%==============================================================================
\section{Expected supernova detection rate}
\label{sec:snrate}
%==============================================================================

% create pickle files with rangePlots/plotRange.py
% plot detection probability with rangePlots/calculateSNDetections.py
\begin{figure}
\includegraphics[width=\linewidth]{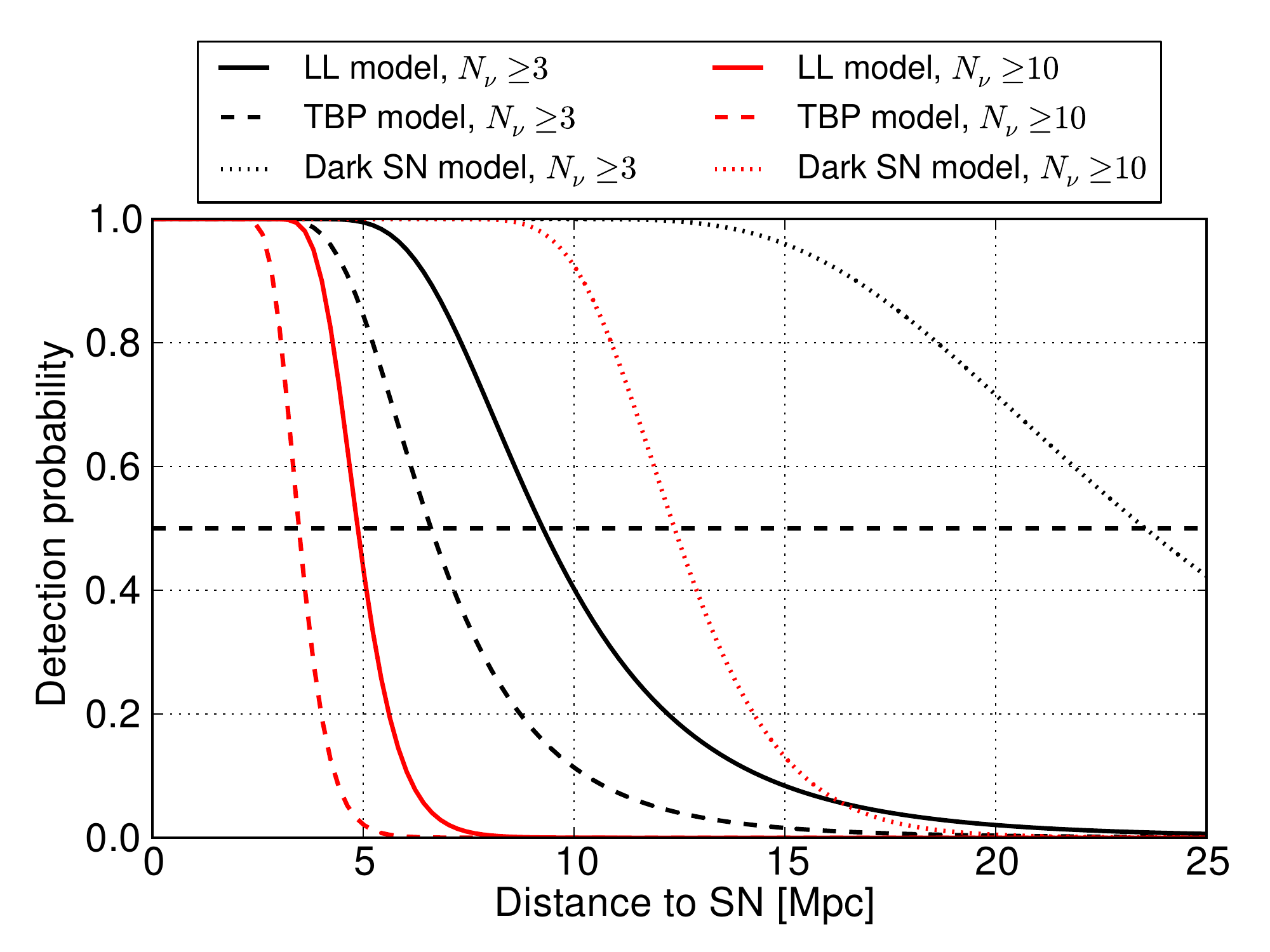}
\caption
  {SN detection probability for the simulated detector after application of
  cuts as described in Section~\ref{sec:background}. Shown are results for
  the LL model, the TBP model, and a dark supernova model (compare
  Table~\ref{tab:sn}) for the detection of at least either 3 or 10 neutrino
  events from the SN.\label{fig:range}}
\end{figure}

Knowing the detector effective area as a function of neutrino energy
(Figure~\ref{fig:backgrounds}\,(c)), we can proceed to calculate the
sensitivity to a supernova neutrino burst with a given spectrum. The
probability to detect a supernova is calculated from Poisson
statistics. We consider a supernova as detected if at least three
neutrino events trigger the detector within 10\un{s}, which is the
threshold we can expect if all other backgrounds can be controlled to
within $1\un{mHz}$.
%If the
%background conditions turn out to be worse than outlined in Section~\ref{sec:background},
%the trigger condition might have to be raised.
Figure~\ref{fig:range} shows the SN detection probability as function of
distance to the SN using the three considered models, with cuts applied against
$\unit[10]{Hz}$ dark noise of the modules and against solar neutrinos (see
Section~\ref{sec:background}). The distance up to which $\geq 3$ neutrinos will
be detected with a probability of $\geq 50\%$ is found to be $6.6\un{Mpc}$ for
the TBP model, $9.3\un{Mpc}$ for the LL model and $25.5\un{Mpc}$ for the dark
SN model, respectively.

With this probability at hand along with the supernova rate in the
local environment, we can compute the rate of expected supernova detections. We
start from a catalog of nearby galaxies that goes up to 100 Mpc~\cite{gwgc}.
Following~\cite{cappellaro} we assume that the blue luminosity of a galaxy is
proportional to the star formation rate and hence also to the supernova rate.
The conversion factor will depend on the galaxy type and is obtained
from SN observations (cf. Table~\ref{tab:sn-rate}). These
conversion factors lead to a total SN rate that is lower by a factor 1.68
compared to a more recent result derived from a comprehensive compilation of
local SNe~\cite{li}. We include this additional scale factor in our rate
estimate to ensure consistency with currently available data. For a limited
number of galaxies, the catalog leaves the type unspecified. Testing the allowed
range of conversion factors, this leads to an error $\leq 4\%$ on the total
core-collapse SN rate for distances up to 20\un{Mpc}.

\begin{table}[tbp]
    \begin{center}
        \begin{tabular}{rlc} \toprule
            {\bf Galaxy} & {\bf Type} & {\bf SN Rate [SNu]} \\
            \midrule
            Elliptical & E-S0 & $<0.05$ \\
            Spiral-like & S0a-Sb & $0.89 \pm 0.33$ \\
            Spiral & Sbc-Sd & $1.68 \pm 0.60$ \\
            Others & Sm, Irr., Pec. & $1.46 \pm 0.71$ \\
            \bottomrule
        \end{tabular}
        \caption[Supernova rates for different galaxy types.]{The expected rates of core-collapse
        supernovae for different galaxy types in supernova units ($\unit[1]{SNu}=\unit[1]{SN}
        (\unit[100]{yr})^{-1} (\unit[10^{10}]{L^B_\odot})^{-1}$). Values
        from~\cite{cappellaro}, scaled by 1.68 (see text).}
        \label{tab:sn-rate}
    \end{center}
\end{table}

\begin{figure}
  \includegraphics[width=\linewidth]{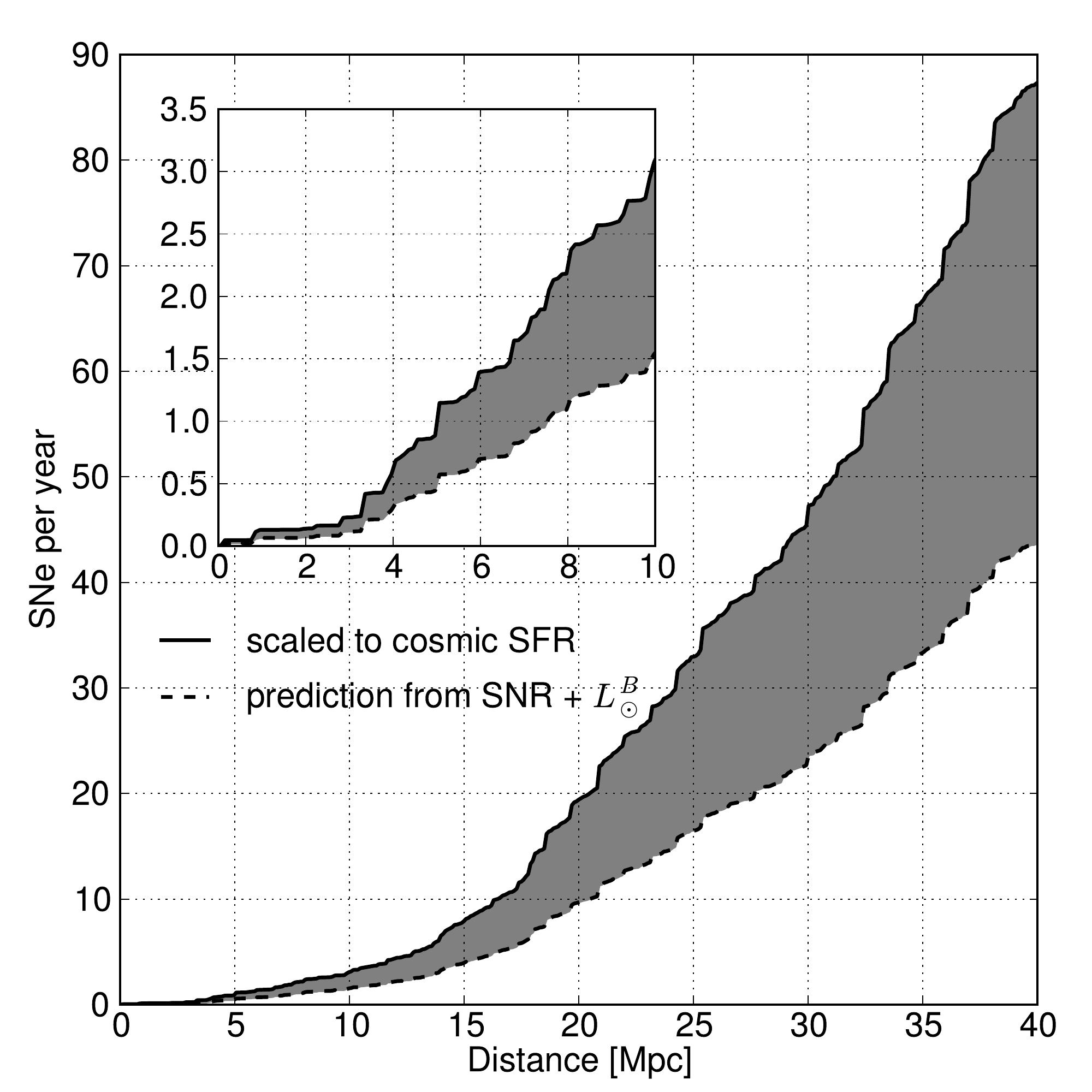}
  \caption[Normalized SN rate and prediction]{Range of cumulative expected
  supernova rate: lower curve (dashed)
  according to observed SN rate and blue luminosity~\cite{cappellaro}, normalized
  to the results of~\cite{li}, upper curve (solid) scaled by a factor 2 to
  match expectations from the cosmic star formation rate~\cite{SNRfromSFR}. The
  galaxy distribution and their blue luminosity values are taken from~\cite{gwgc}.}
  \label{fig:snrate-40mpc}
\end{figure}

Comparing with a theoretical prediction based on the initial mass function and
cosmological star formation rate~\cite{SNRfromSFR}, our SN rate is still a
factor two lower. One explanation might be that the conversion factors in
Table~\ref{tab:sn-rate} are based on observations with a bias to miss many faint
SNe. Another possibility is a significant contribution of dark supernovae, that
would not be detected optically, but still emit neutrinos~\cite{SNRfromSFR}.
Scaling our blue luminosity prediction by a factor of two, we also find good
agreement with the observed rate of nearby SNe~\cite{Horiuchi:2013fk}, and
regard the rate estimates based on actual observations of SNe and those
scaled to the star formation rate---both shown in Figure~\ref{fig:snrate-40mpc}---as lower and upper limit, respectively.

Table~\ref{tab:detections} gives a summary of expected SN detections per decade
in different neutrino event multiplicity bins. The total resulting number of SN
detections with the cuts described in Section~\ref{sec:background} ranges from
20 to 41 per decade for the LL model and about half of that for the TBP model.
For dark SNe, we make the assumption that SNe collapse to a black hole
at a rate of $10\%$ of the regular core-collapse SN rate. Yet due to a more
energetic neutrino spectrum, dark SNe are detected at a higher efficiency
resulting in between 17 and 35 dark observations per decade, comparable to the
number of detections from the LL model. Altogether, one can expect to observe
at least one SN per year on average, perhaps up to 5 or more. The rate of SNe
producing strong bursts of ten or more neutrinos is between 0.2 and 0.5 per
year without dark SNe, reaching up to almost 1 event per year if dark SNe are
included. Note that while for a neutrino multiplicity of three, the event is as
likely to be from a SN as from detector noise or solar background, a single event with a
multiplicity of four (five) already constitutes a SN signal of $\sim 2\sigma$
($\sim 4\sigma$). Yet, the true power of the approach lies in the combination
with follow-up missions, that can detect the same SN in the optical or X-ray
regime.

\begin{table}[tbp]
    \begin{center}
        \begin{tabular}{lccccc}
            \toprule
            &                  & \multicolumn{2}{c}{{\bf CCSNe}} & {\bf CCSNe}  & {\bf Dark} \\
            & $\mathbf{N_\nu}$ & \multicolumn{2}{c}{{\bf (LL)}}  & {\bf (TBP)}    & {\bf SNe} \\
            &                  & {\bf deep} & {\bf shallow}      & {\bf deep}     & {\bf deep} \\
            \midrule
            \multirow{6}{*}{\parbox[c]{3ex}{$\mathbf{L_\odot^{B}}$}}  & $\geqslant 3$  & 20.4 & 3.1 & 10.3 & 17.4 \\
                                                                      & $\geqslant 4$  & 13.0 & 6.0 & 7.0 & 11.9 \\
                                                                      & $\geqslant 5$  & 9.9 & 7.4 & 5.4 & 8.4 \\
                                                                      & $\geqslant 6$  & 8.2 & 7.0 & 4.4 & 6.3 \\
                                                                      & $\geqslant 7$  & 7.0 & 6.0 & 3.7 & 4.8 \\
                                                                      & $\geqslant 10$ & 4.9 & - & 2.3 & 2.7 \\
            \midrule
            \multirow{2}{*}{\parbox[c]{3ex}{\bf SFR}} & $\geqslant 3$  & 40.9 & 6.1 & 20.6 & 34.8 \\
                                                      & $\geqslant 10$ & 9.8 & - & 4.5 & 5.4 \\
            \bottomrule
        \end{tabular}
        \caption[SN detection rate per decade]{Expected number of supernova
        detections within one decade based on SN rates computed from the blue
        luminosity $L_\odot^{B}$ of galaxies~\cite{li,cappellaro} (first 6
        lines). The last two lines are for a prediction scaled to match the
        star formation rate (SFR)~\cite{SNRfromSFR}
        (c.f.~Figure~\ref{fig:snrate-40mpc}). Dark SNe are assumed to occur at
        a fraction of $10\%$ of all core-collapse SNe (CCSNe). For the LL
        model, values for the simulated shallow ice detector (see
        Section~\ref{sec:diffusive}) are listed as well.
        }
        \label{tab:detections}
    \end{center}
\end{table}

%==============================================================================
\section{Conclusion}
\label{sec:conclude}
%==============================================================================

A $\unit[{\cal O}(10)]{Mton}$ scale neutrino detector is required to extend
the sensitivity to core-collapse SNe beyond the Large Magellanic Cloud
to neighboring galaxies and yield a detection rate of up to several
SNe per year. In this paper, we have explored an implementation of such
%two implementations for
%such a detector in the Antarctic ice: either in the diffusive ice
%(around $900\un{m}$ below surface) where photon transport is diffusion
%dominated or in very clear ice (at around $2300\un{m}$ depth) where
%photons can travel with very little scattering. In both cases, the
%detector geometry would resemble that of IceCube, however, with a much
%reduced string spacing and significantly increased density of
%photo-sensors per string.
a detector in the very clear Antarctic ice at around $\unit[2300]{m}$
depth below surface where photons can travel with very little
scattering. The detector geometry would resemble that of IceCube,
however, with a much reduced string spacing and significantly
increased density of photo-sensors per meter.  Such a detector could
be built in a similar manner as IceCube, by drilling holes into the
ice and deploying strings holding the light sensors.  To achieve the
$\unit[{\cal O}(10)]{Mton}$ detector with sensitivity to 10 MeV
neutrinos at trigger level would require a
%61-string installation with about 20,000 photo-sensors similar to the
%HQE sensors employed by IceCube for the diffusive ice location and
%about ten times that total photo-cathode area for the clear ice
%location.
127-string installation with about 50 times the total photocathode
area used in IceCube, indicating the demand for new and cheaper
technology and dedicated R\&D for photo-sensors with large effective
photo-cathode area. Work in this direction has already been initiated
and is embedded into the IceCube low-energy extension project PINGU
(see \cite{PINGU-LoI}, section 14). Efforts include the use of wavelength-shifters
\cite{WOM_ICRC} and of multiple small PMTs within a single module
similar to the KM3NeT optical modules \cite{KM3NeT-OM}.
Besides the large effective area, we identify a low noise rate as
crucial requirement for the sensors.
%Even for a self-noise rate of
%10\un{Hz} per meter instrumented string, the spatial and temporal
%distribution of hits has to be exploited and the number of strings
%increased to 127 to retain the ${\cal O}(10)$ megaton target effective
%mass.
Above a self-noise rate of $\unit[\sim50]{Hz}$ per meter instrumented
string, it becomes difficult to retain the ${\cal O}(10)$ megaton
target effective mass when exploitation the spatial and temporal
distribution of hits to suppress sensor self-noise.
%The short scattering length will render this method void for the
%diffusive ice, making the task to separate neutrino events from random
%noise triggers yet more difficult.

The main physical background remaining after the self-noise reduction arises
due to solar $^8$B neutrinos. These can be identified via lower photon
multiplicity or---contingent on the ability to reconstruct their direction---by
their angular proximity to the Sun.  Atmospheric muons will provide exceedingly
bright events in such a dense installation, resulting in a small downtime for
the deep ice. More challenging is the rejection of Michel electron events, yet
with the IceCube detector fully surrounding the array, their identification
will be straight-forward and only a modest performance reduction will arise
from their rejection. Muon induced spallation events need to be suppressed using
temporal and spatial correlations with the originating muon and require more
detailed studies including accurate description of the nuclear processes as
well as full event reconstruction, which are both beyond the scope of this
paper.

Using a catalog of nearby galaxies, we have computed the rate of
detectable SN neutrino bursts. We show that depending on the SN rate
and explosion model assumed, we can expect to observe between 10 and
41 SNe per decade in neutrinos (not counting the dark SNe). By
combining the observation of a neutrino burst with an optical
detection of a SN, one can thereby disentangle the question of
explosion model and SN rate. We note that in the future, with large,
wide field optical surveys covering essentially the full sky, nearby
SNe should be found in an even more systematic manner than today
\cite{LSST,ZTF}.

Dark supernovae, where the star collapses to a black hole, could
produce more than 20 detectable bursts per decade. Such a SN can be
indirectly identified through the absence of an optical counterpart
(with the risk of confusing it with a regular, dust obscured SN), or
more directly, by observing neutrinos of higher energies. The limited
energy resolution of the detector ($\approx 30\%$ per neutrino event)
should be sufficient to estimate the effective temperature of the
neutrino emission and hence the origin of the burst.

A neutrino detector as described in this paper will not only yield a
precise measurement of the local supernova rate and can uncover dark
supernovae. A few of the supernovae will be closer, or perhaps even
galactic, and hence yield a much higher number of coincident
neutrinos, allowing to infer details about the explosion or even the
neutrino mass hierarchy~\cite{Dighe:2003be,Serpico:2011ir}. To conclude, we point out that the low energy
threshold and very large effective mass make such a detector
potentially interesting for a number of other physics phenomena,
including e.g.\ proton decay studies and solar neutrino analysis.
While we have demonstrated that one can achieve the desired
goal---routine observation of SNe in neutrinos---we acknowledge the
assumptions we have made on our way.  In particular, we have
identified the development of large-area low-noise photo-sensors,
suitable for deployment in the ice at South Pole, as a key requirement.

\section*{Acknowledgements}
We would like to thank Imre Bartos, John Beacom, Francis Halzen, Allan
Hallgren, and Lutz K\"opke for fruitful discussions, as well as the
IceCube collaboration for graciously allowing us to use its photon
tracking software. We acknowledge support from the Helmholtz Alliance
for Astroparticle Physics (HAP).

%%% appendix sections are then done as normal sections
%\appendix
%
%\section{Nora's SN detection rate}
%    \begin{center}
%        \begin{tabular}{rcccccc} \toprule
%            % Nora:
%            Nora & \multicolumn{2}{c}{{\bf CC SN (LL)}} & \multicolumn{2}{c}{{\bf CC SN (TBP)}} &\multicolumn{2}{c}{{\bf Dark SN}}\\
%            {\bf detectable neutrinos} & diffusive & clear & diffusive & clear & diffusive & clear\\ \midrule
%            3 & $6.1$& $4.9$ & $4.5$ & $3.3$ & $5.4$ & $6.3$ \\
%            4 - 5 & $3.7$ & $3.0$ & $2.7$ & $2.2$ & $3.8$ & $4.5$ \\
%            6 - 9 & $2.0$ & $1.8$ & $1.8$ & $1.5$ & $2.0$ & $2.5$ \\
%            $\geqslant 10$ & $2.7$ & $2.3$ & $2.0$ & $1.6$ & $1.5$ & $1.7$\\ \midrule
%            {\bf total} & {\bf 15-29} & {\bf 12-24} & {\bf 11-22} & {\bf 9-18} & {\bf13-26} & {\bf15-30} \\
%        \end{tabular}
%%        \caption{Compare Table \ref{tab:detections}. The expected number of supernova detections within
%%        one decade according to the observed supernova rate (compare Figure \ref{fig:snrate-40mpc}). Dark
%%        supernovae are assumed to contribute $10\%$ to the total supernova rate. No correction for neutrino
%%        detection efficiencies has been taken into account.}
%    \end{center}


%% References
%% Usage of~\cite is as follows:
%%~\cite{key}         ==>>  [#]
%%~\cite[chap. 2]{key} ==>> [#, chap. 2]
%%


%\section*{References}

%\bibliographystyle{model1a-num-names}
\bibliographystyle{elsarticle-num}
\bibliography{bibdatabase}






% \end{thebibliography}

\end{document}